\documentclass[a4paper,fleqn,usenatbib]{mnras}

\usepackage{newtxtext,newtxmath}

\usepackage[T1]{fontenc}
\usepackage{ae,aecompl}

\usepackage{graphicx}	% Including figure files
\usepackage{amsmath}	% Advanced maths commands
\usepackage{amssymb}	% Extra maths symbols
\usepackage{bm}
\usepackage{caption}
\usepackage{multicol}
\usepackage{hyperref}
\usepackage{booktabs}
\usepackage{ulem}

\def\eqref#1{equation~(\ref{#1})}

\def\Figref#1{Fig.~\ref{#1}}

\newcommand {\Myr}{\,{\rm Myr}}
\newcommand {\Gyr}{\,{\rm Gyr}}

\newcommand {\kpc}{\,{\rm kpc}}
\newcommand {\kms}{\,{\rm km}\,{\rm s}^{-1}}

\newcommand {\kpckms}{\,{\rm kpc}\,{\rm km}\,{\rm s}^{-1}}
\newcommand {\kmskpc}{\,{\rm km}\,{\rm s}^{-1}\,{\rm kpc}^{-1}}

\newcommand {\phisun}{{\varphi_\odot}}

\newcommand {\vRsun}{v_{R\odot}}
\newcommand {\vphisun}{v_{\varphi\odot}}

\newcommand {\vx}{{\bm x}}
\newcommand {\vvel}{{\bm v}}

\newcommand {\Rg}{R_{\rm g}}
\newcommand {\Ra}{R_{\rm a}}
\newcommand {\Rb}{R_{\rm b}}

\newcommand {\vR}{v_R}
\newcommand {\ovR}{\bar{v}_R}
\newcommand {\vphi}{v_\varphi}
\newcommand {\vphimax}{v_{\varphi,{\rm max}}}
\newcommand {\vphimin}{v_{\varphi,{\rm min}}}

\newcommand {\pR}{p_R}
\newcommand {\pphi}{p_\varphi}
\newcommand {\RCR}{R_{\rm CR}}
\newcommand {\RCRo}{R_{\rm CR0}}
\newcommand {\vCR}{v_{\rm CR}}
\newcommand {\Lz}{L_z}
\newcommand {\Lc}{L_{\rm c}}
\newcommand {\Lzo}{L_{z0}}
\newcommand {\phib}{\varphi_{\rm b}}
\newcommand {\phibar}{\varphi_{\rm b}}
\newcommand {\sigpar}{{\sigma_{\rm p}}}
\newcommand {\vc}{v_{\rm c}}
\newcommand {\Uh}{U}

\newcommand {\vJ}{{\bm J}}
\newcommand {\JR}{J_R}

\newcommand {\dJR}{\dot{J}_R}
\newcommand {\Jphi}{J_\varphi}
\newcommand {\dJphi}{\dot{J}_\varphi}

\newcommand {\Jf}{J_{\rm f}}
\newcommand {\Js}{J_{\rm s}}
\newcommand {\dJf}{\dot{J}_\mathrm{f}}
\newcommand {\dJs}{\dot{J}_\mathrm{s}}
\newcommand {\vJres}{{\bm J}_{\rm res}}
\newcommand {\Jsres}{J_{\rm s, res}}

\newcommand {\DJsmax}{\Delta J_{\rm s, max}}
\newcommand {\Jl}{J_\ell}
\newcommand {\Jlsep}{J_{\ell,{\rm sep}}}
\newcommand {\vtheta}{{\bm \theta}}
\newcommand {\thetaR}{\theta_R}
\newcommand {\thetaphi}{\theta_\varphi}
\newcommand {\thetas}{\theta_{\rm s}}
\newcommand {\dthetas}{\dot{\theta}_{\rm s}}

\newcommand {\thetaf}{\theta_{\rm f}}
\newcommand {\dtheta}{\dot{\theta}}
\newcommand {\ddtheta}{\ddot{\theta}}
\newcommand {\thetares}{\theta_{\rm res}}
\newcommand {\thetasep}{\theta_{\rm sep}}

\newcommand {\Omegap}{\Omega_{\rm p}}
\newcommand {\Omegapo}{\Omega_{\rm p0}}
\newcommand {\dOmegap}{\dot{\Omega}_{\rm p}}
\newcommand {\ddOmegap}{\ddot{\Omega}_{\rm p}}
\newcommand {\OmegaR}{\Omega_R}

\newcommand {\Omegaphi}{\Omega_\varphi}

\newcommand {\vOmega}{\bm \Omega}
\newcommand {\Omegas}{\Omega_s}
\newcommand {\Phio}{\Phi_0}
\newcommand {\Phib}{\Phi_{\rm b}}
\newcommand {\Phim}{\Phi_m}
\newcommand {\Psik}{\Psi_{\bm k}}
\newcommand {\vN}{{\bm N}}
\newcommand {\vk}{{\bm k}}

\newcommand {\NR}{N_R}
\newcommand {\Nphi}{N_\varphi}
\newcommand {\kf}{k_{\rm f}}
\newcommand {\ks}{k_{\rm s}}
\newcommand {\e}{\mathrm{e}}
\newcommand {\bH}{\bar{H}}
\newcommand {\Ep}{E_{\rm p}}
\newcommand {\Epsep}{E_{\rm {p,sep}}}
\newcommand {\EJ}{E_{\rm J}}
\newcommand {\Af}{A_{\rm f}}
\newcommand {\tres}{t_{\rm res}}
\newcommand {\Tl}{T_\ell}
\newcommand {\Tres}{T_{\rm res}}

\title[]{Resonance sweeping by a decelerating Galactic bar}

\author[R. Chiba et al.]{
Rimpei Chiba,$^{1}$\thanks{E-mail: rimpei.chiba@physics.ox.ac.uk}
Jennifer K. S. Friske,$^{2}$
and Ralph Sch{\"o}nrich$^{3}$
\\
$^{1}$University of Oxford, Rudolf Peierls Centre for Theoretical Physics, OX1 3PU Oxford, UK\\
$^{2}$Ludwig-Maximilians-Universit{\"a}t, Fakult{\"a}t f{\"u}r Physik Schellingstr. 4, 80799 M{\"u}nchen, Germany\\
$^{3}$Mullard Space Science Laboratory, University College London, Holmbury St. Mary, Dorking, Surrey, RH5 6NT, UK
}

\date{Accepted XXX. Received YYY; in original form ZZZ}

\pubyear{2020}

\begin{document}
\label{firstpage}
\pagerange{\pageref{firstpage}--\pageref{lastpage}}
\maketitle


\begin{abstract}

We provide the first quantitative evidence for the deceleration of the Galactic bar from local stellar kinematics in agreement with dynamical friction by a typical dark matter halo. The kinematic response of the stellar disk to a decelerating bar is studied using secular perturbation theory and test particle simulations. We show that the velocity distribution at any point in the disk affected by a naturally slowing bar is qualitatively different from that perturbed by a steadily rotating bar with the same current pattern speed $\Omegap$ and amplitude. When the bar slows down, its resonances sweep through phase space, trapping and dragging along a portion of previously free orbits. This enhances occupation on resonances, but also changes the distribution of stars within the resonance. Due to the accumulation of orbits near the boundary of the resonance, the decelerating bar model reproduces with its corotation resonance the offset and strength of the Hercules stream in the local $v_R$-$v_\varphi$ plane and the double-peaked structure of mean $\vR$ in the $\Lz$-$\varphi$ plane. At resonances other than the corotation, resonant dragging by a slowing bar is associated with a continuing increase in radial action, leading to multiple resonance ridges in the action plane as identified in the \textit{Gaia} data. This work shows models using a constant bar pattern speed likely lead to qualitatively wrong conclusions. Most importantly we provide a quantitative estimate of the current slowing rate of the bar $\dOmegap = (-4.5 \pm 1.4) \kms\kpc^{-1}\Gyr^{-1}$ with additional systematic uncertainty arising from unmodeled impacts of e.g. spiral arms.

\end{abstract}


%%%%%%%%%%%%%%%%%%%%%%%%%%%%%%%%%%%%%%%%%%%%%%%%%%

%%%%%%%%%%%%%%%%% BODY OF PAPER %%%%%%%%%%%%%%%%%%

\section{Introduction}
\label{sec:introduction}

\subsection{Slowing bar as probe for dark halo kinematics}
\label{sec:first_introduction}

It is widely accepted that our Galaxy has a prominent, rotating stellar bar at its centre, as do roughly half of known disc galaxies \citep[]{sellwood1993dynamics,Buta2015Classical}. Bars cannot be statically rotating objects, since they form part of a delicate angular momentum balance: loss to dark halo and stellar disc, and gains from funneling gas to small radii. Analytical models and simulations of the Galactic bar in the presence of a dark matter predict that a bar experience angular momentum loss due to dynamical friction, slowing their rotation frequency, the so called pattern speed $\Omegap$, and hence making them grow \citep[][]{weinberg1985evolution,hernquist1992bar,debattista2000constraints,valenzuela2003secular,Martinez2006Evolution}. This angular momentum loss is proportional to the density of the dark matter halo, but also depends strongly on the velocity distribution of the dark matter \citep[][]{athanassoula2003determines}. The amount of angular momentum transfer would also be drastically altered e.g. with modified theories of gravity (requiring different amounts of dark matter), or if the dark halo is in the form of a degenerate quantum condensate \citep[e.g.][]{Goodman2000Repulsive,Bohmer2007BEC}. On the other hand, bars gain large amounts of angular momentum, offsetting some fraction of the above loss, by funneling gas towards the Galactic centre \citep[][]{Albada82,Regan2004BarDriven}, where the gas feeds the central black hole and gets expelled by it \citep[][]{Silk98, BlandHawthorn03}, and/or forms a massive nuclear disc, as found in the Milky Way \citep[][]{Launhardt02, Schoenrich15}.

While the density distribution of the dark matter halo can be mapped from its gravitational potential \citep[e.g.][]{Iocco2011Dark,Cole2017centrally} by comparing the total mass of the Galaxy, inferred from the Galactic rotation curve \citep[e.g.][]{Sofue2013Rotation}, with the baryonic mass inferred e.g. from infrared maps \citep[e.g.][]{Robin2012Stellar}, interstellar gas maps \citep[e.g.][]{Nakanishi2006ISM}, and gravitational microlensing \citep[e.g.][]{Alcock1995Microlensing}, the detailed kinematic state of the dark matter is only accessible by dynamical modeling, making (if measured) the slowing rate of the bar $\dOmegap$ an important constraint for the nature of the dark matter.

Despite this importance, and despite the theoretical requirement that bars have to be strongly evolving, there is yet no observational evidence of a slowing bar. Few papers, however, have predicted indirect signatures of an evolving bar: \cite{weinberg1994kinematic} studied the orbits trapped in resonance of a slowing bar and showed that the deceleration results in an increased velocity dispersion near the outer Lindblad resonance. However, both his model and the available data back then was not detailed enough to draw firm conclusions on the bar slow-down. \cite{aumer2015origin} linked the slowing/growth of the bar to the discovery of high line-of-sight velocity tails observed in the distribution of stars within the bar \citep{Nidever12}. However, this signal strongly depends on the subsequent diffusion out of these orbits and the surrounding disc kinematics, so is unlikely to yield a measurement of $\dOmegap$. \cite{Halle2018Radial} showed in N-body simulations that stars trapped in the co-rotation resonance of the bar are churned radially outwards by the slowing bar and \cite{Khoperskov2019Escapees} recently linked this mechanism to the high-metallicity stars in the Solar vicinity \citep{Grenon1999Kinematics}. Investigating the correlation between the kinematics and chemistry of stellar disk could be a promising method to prove the bar slow-down, since trapped stars are expected to have smaller birth radius and thus higher metallicities.

Our best bet to trace the bar slow-down are stellar kinematics in the Solar neighbourhood with full 6D phase space information. Motions of local stars are known to be strongly affected by the bar's gravitational perturbation, especially near resonances, i.e. when stellar orbital frequencies are in commensurable relation with the bar's pattern speed. As the bar decelerates, the resonance regions sweep through the stellar phase-space, trapping and dragging a number of orbits, leaving noticeable changes in the stellar distribution. Therefore the current local kinematics can be used as archaeological evidence to probe the evolutionary history of the bar \citep[]{weinberg1994kinematic}. However, most past studies attempting to fit the stellar streams in the Solar neighbourhood use a static bar with constant $\Omegap$ \citep[e.g.][]{dehnen2000effect}. One exception is a notion of a suddenly formed, very young bar (albeit still with a constant $\Omegap$), which would leave some transient effects lingering in stellar kinematics \citep{minchev2010low}. However, such a young age appears not fully in line with the low relative star-formation rates in the Milky Way's nuclear disc. \cite{fux2001order} attributed little importance to the effects of a slowing bar, claiming that it mainly introduces a delayed response.

In contrast, this paper will show the importance of a slowing bar for resonances in the Milky Way disc, in particular for the local kinematic substructures observed by \textit{Gaia}. We use secular perturbation theory and test particle simulations to explore how resonances of a slowing bar capture and drag orbits. In the next subsection, we summarise the observed kinematic substructure in the Solar neighborhood, which has been used in the past to judge the pattern speed and strength of the Galactic bar. We will also provide a first glimpse of how much a slowing bar model differs from the predictions of a constant $\Omegap$ model with otherwise identical parameters. We will list the main observable features explained by the slowing bar model, which in turn provide us with the leverage to estimate the current $\dOmegap$. At the end of Section \ref{sec:observation}, we give an outline of the paper.

\begin{figure*}
  \begin{center}
    \includegraphics[width=18cm]{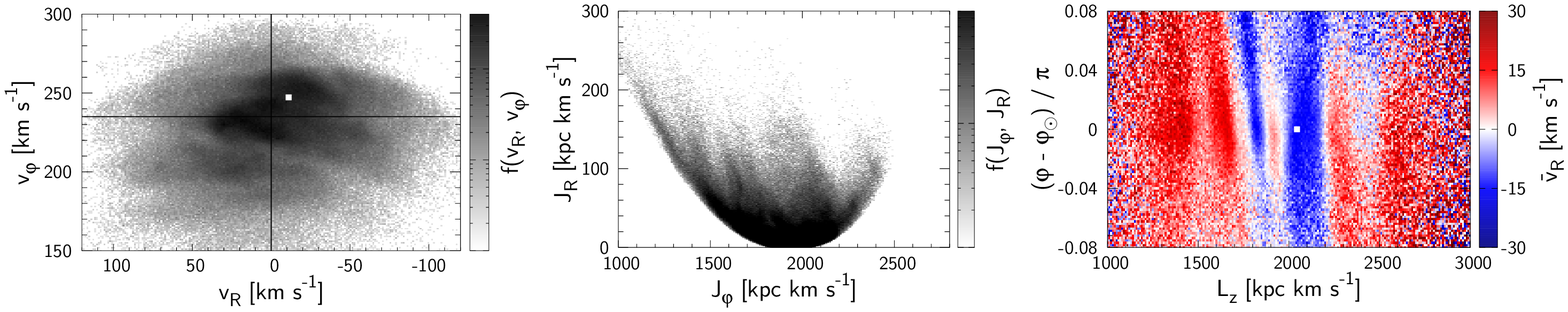}
    (a)~Local kinematics observed by \textit{Gaia} DR2.
    \includegraphics[width=18cm]{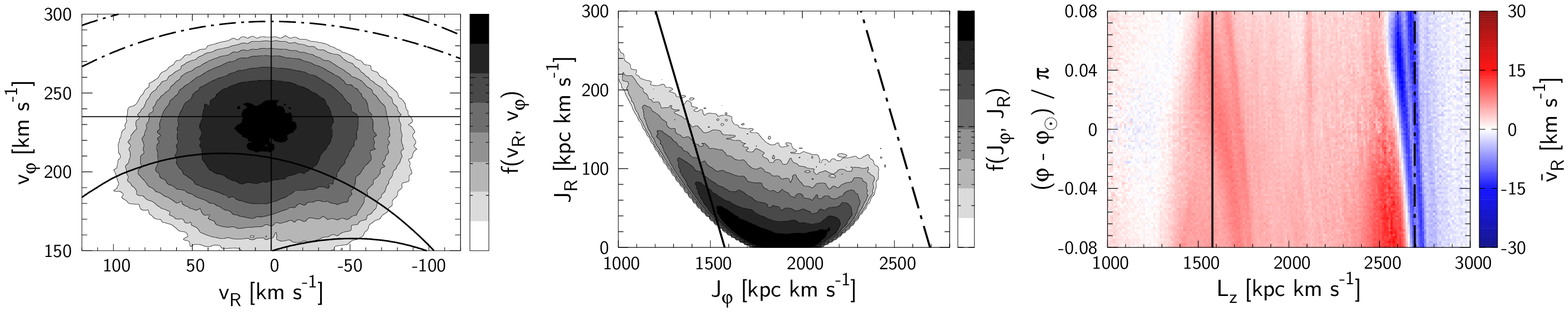}
    (b)~Test-particle simulation with a steadily rotating bar.
    \includegraphics[width=18cm]{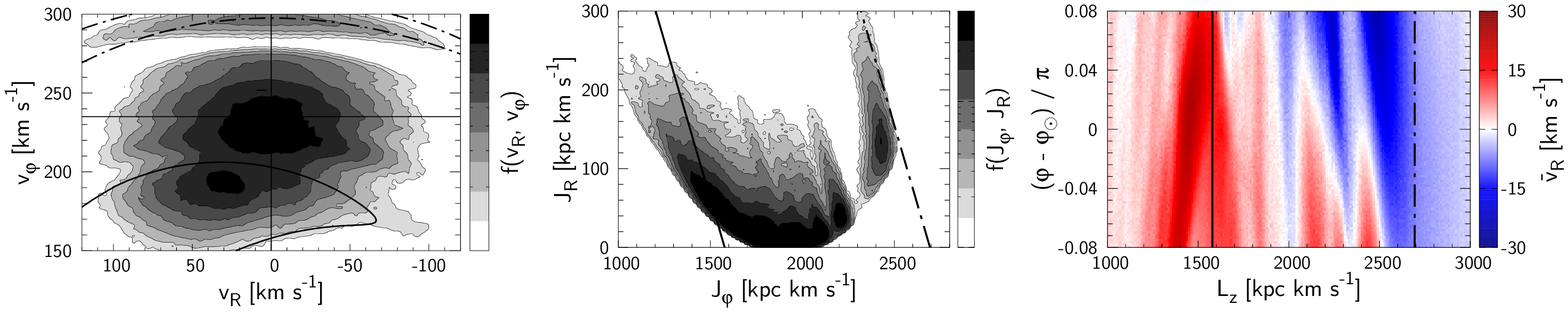}
    (c)~Test-particle simulation with a slowing bar.
  \caption{Top panel: Kinematic data from \textit{Gaia} DR2 with quality cut on parallax of $p/\sigpar > 5$. For the distribution in velocity space (left) and action space (middle), we restrict the samples to heliocentric distance $s < 0.3 \kpc$. The small white square indicates the coordinate of the Sun (Section \ref{sec:coordinate}). Middle row: Test particle simulation of a constantly rotating quadrupole bar with pattern speed $\Omegap = 35 \kmskpc$. Solid lines mark the corotation resonance, dashed lines the outer Lindblad resonance. Bottom row: Test particle simulation of a \textit{rapidly} slowing bar presented in \ref{sec:rapidly_slowing_bar}. Snapshot taken at $\Omegap = 35 \kmskpc$. Apart from the deceleration, the bar parameters are identical to the constant pattern speed model.}
  \label{fig:GaiaDR2}
  \end{center}
\end{figure*}

\subsection{Kinematic structure of the Solar neighborhood}
\label{sec:observation}

The top row of \Figref{fig:GaiaDR2} shows the kinematic substructure revealed by \textit{Gaia} DR2 \citep[][]{GaiaDR2GaiaCollaboration,GaiaDR2Katz2019} with parallax offset and distance derivation from \cite[][]{Schoenrich2019distance}. We identify the 2D in-plane structure in three different statistics:

\begin{itemize} 
  \item Left-hand panel: Velocity distribution $f(\vR,\vphi)$ of local stars showing substructures which have long been suspected to be caused by resonances with non-axisymmetric components of the Galaxy \citep[e.g.][]{kalnajs1991pattern}. In particular, the large sub-population seen at low $\vphi$ and positive $\vR$, known as the Hercules stream, has been extensively modeled with bar resonances \citep[e.g.][]{dehnen2000effect}.
  \item Middle panel: Distribution in the action plane $f(\Jphi,\JR)$ estimated in an axisymmetric logarithmic potential, which has been identified by \cite{Sellwood10} (using data from the Geneva Copenhagen survey) to show structures along the resonant lines and more recently been used by works on \textit{Gaia} DR2 \citep[e.g.][]{Trick2019Identifying}.
  \item Right-hand panel: Mean radial velocity $\ovR$ plotted over the angular momentum $\Lz$ and Galactic azimuth $\varphi$. \textit{Gaia}'s large-scale coverage permitted the first probe into the spatial dependence of the kinematic structure. Each stripe shows a different azimuthal dependence, indicating a distinct origin \citep[]{friske2019more}.
\end{itemize}

With this phase space information, one should in principle be able to identify the positions of resonances with the bar and thus predict the bar's pattern speed. Yet this task is plagued by degeneracies; there are currently many possible models which can reproduce the observed features with different resonances. This has led to a major debate between proponents of a fast/short bar ($\Omegap \gtrsim 50 \kmskpc$) and a slow/long bar ($\Omegap \lesssim 40 \kmskpc$), where the debate has mainly concentrated on the cause of the Hercules stream: Fast bar proponents \citep[e.g.][]{dehnen2000effect, antoja2014constraints,fragkoudi2019ridges} interpreted the Hercules stream as stars near the outer Lindblad resonance (OLR), which well matches the strength of the feature and the offset in $\vR$, though the required high pattern speed contradicts with the modeling of the bar/bulge using red clump stars \citep[]{Portail2017Dynamical} and studies on the inner gas dynamics \citep[]{sormani2015gas3}. Slow bar proponents \citep[][]{perez2017revisiting, Monari2019signatures, binney2017orbital, DOnghia2020Trojans} interpreted the Hercules stream as due to orbits trapped in the corotation resonance (CR). However, models with constant slow pattern speeds tend to underpredict the strength of the observed feature (or vice versa require a too strong bar).

In the middle row of \Figref{fig:GaiaDR2}, we present a test particle simulation for such a slow quadrupole bar rotating with a constant $\Omegap = 35 \kmskpc$ and reasonable strength fitted to the model of \cite{sormani2015gas3} (see the main text for details). We mark the CR and the OLR in solid and dot-dashed lines. As mentioned above, the Hercules stream is underpredicted in the velocity plane (left) as reported by many other authors. The bottom row displays our slowing/elongating bar model with otherwise identical parameters. The deceleration of the bar increases the strength of the Hercules stream, as well as offsetting it towards larger $\vR$, in better agreement with the data. This difference is also indicated in N-body studies: with a steadily rotating bar, \cite{fragkoudi2019ridges} reported that the CR do not create a prominent feature in the Solar neighbourhood, while with a self-consistently slowing bar, \cite{DOnghia2020Trojans} confirmed a clear asymmetry in $\vR$ at the CR akin to the Hercules stream. Our slowing bar model also produces strong resonance features of high order resonances in between the CR and the OLR as confirmed in the action plane (middle). We can also see in the $\ovR(\Lz,\varphi)$ plane (right) that the CR appear as a spear-head structure which is identifiable in the \textit{Gaia} data near $\Lz \sim 1500 \kpckms$. 

We note, however, that perturbation by yet unconstrained spiral arms offers additional freedom in reproducing the data: \cite{Hunt2018Transient} showed that repeated perturbation by transient winding spiral arms, ubiquitously seen in N-body simulations, can reproduce the Hercules stream either with or without the presence of a bar, and \cite{Sellwood2019Discriminating} showed how some of the structures in action-angle space can be linked with spiral arms. We also note that our paper employs the simplest possible model for a slowing bar; we did not include higher-order modes of the bar which will enhance the signature of minor resonances \citep[]{Monari2019signatures}. Therefore, we do not aim at a model that accounts for all observed features, but rather to show the significant impact of the deceleration of the bar.

The paper is organized as follows: In Section \ref{sec:theory}, we introduce our slowing bar model and discuss resonance dragging/drift in actions using secular perturbation theory. Section \ref{sec:test_particle_sim} describes the method of our test-particle simulation. In Section \ref{sec:results} we start our discussion with a constantly rotating bar and subsequently explore the kinematic consequence of a slowing bar. Section \ref{sec:conclusion} concludes.

\section{Theory}
\label{sec:theory}

\subsection{Coordinate frame}
\label{sec:coordinate}

Throughout the paper, we take the position of an observer at the Galactic South Pole, thus using positive pattern speed, azimuthal velocity and angular momentum. In our frame the radial velocity $\vR$ points outwards in contrast to the usual heliocentric radial velocity $\Uh$. We use Galactic circular speed $\vc = 235 \kms$ \citep{Reid2019Trigonometric}, Solar galactocentric radius $R_0 = 8.2 \kpc$ \citep{Gravity2019geometric}, $\phib - \phisun = 30^\circ$ for the Solar galactocentric azimuthal angle with respect to the bar major axis $\phib$ \citep{wegg2015structure}, and Solar velocity $(\vRsun, \vphisun - \vc) = (-11.1, 12.24) \kms$ in concordance with previous findings \citep[][]{Joshi07, SBD, S12, McMillan17}. Since we deal with a slowing bar, we work in an inertial frame to make explicit the time dependence of $\Omegap(t)$.

\subsection{Model}
\label{sec:model}

We study orbits perturbed purely by a slowing bar. We thus neglect self-gravitational effects in our model and assume a logarithmic background potential corresponding to a constant circular speed $\vc$. We further simplify the discussion by restricting the model to 2D in-plane motion, and by modeling the bar as a $m = 2$ quadrupole, rotating like a rigid body (i.e. no flexing or winding up):
\begin{align}
  \Phi(R, \varphi, t) 
  &= \Phio(R) + \Phib(R, \varphi, t) \nonumber \\
  &= \vc^2 \ln \left(R\right) + \Phim(R, t) \cos m\left[\varphi - \int_0^t dt'~\Omegap(t') \right],
  \label{eq:potential_polar}
\end{align}
where $\Omegap(t)$ denotes the time-dependent pattern speed and $\phib = \int_0^t dt' \Omegap(t')$ expresses the current azimuth of the bar major axis (we choose $\Phim < 0$). In this paper $m$ always refers to 2, although we retain the expression $m$ to keep our discussion general and to avoid confusion with other factors of 2. Studies from N-body simulations \citep[e.g.][]{aumer2015origin} imply that the bar's slowing rate, $-\dOmegap$, decreases with time ($\ddOmegap > 0$). A reasonable model for the pattern speed is thus $\Omegap(t) \propto t^{n}$, where $n < 0$. In approximation to \cite{aumer2015origin}, we choose $n = -1$, which corresponds to a linear increase in co-rotation radius $\RCR$. Therefore
\begin{equation}
  \Omegap(t) = \frac{\vc}{\RCR(t)} = \frac{\vc}{\RCR(0) + \vCR t},
  \label{eq:pattern_speed}
\end{equation}
where $\vCR$ is the velocity of the co-rotation radius. The bar's slowing rate is best described with the following dimensionless parameter:
\begin{equation}
  \eta \equiv - \frac{\dOmegap}{\Omegap^2} = {\rm const}.
  \label{eq:slowing_rate}
\end{equation}
Since we only consider the case where the bar is slowing down ($\dOmegap < 0$), $\eta$ is defined to be positive. In our model with a flat rotation curve, $\eta$ is $\vCR/\vc$, the dimensionless representation of $\vCR$.

\begin{figure}
  \begin{center}
    \includegraphics[width=8.9cm]{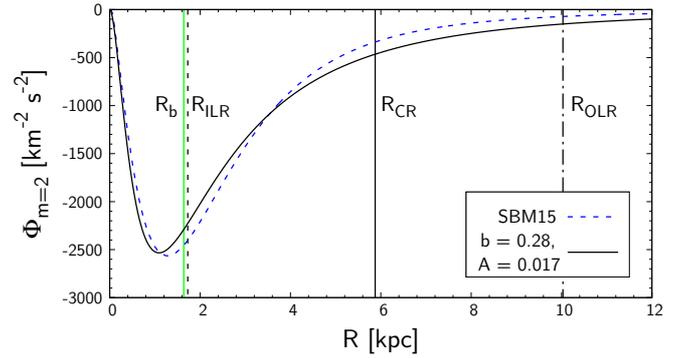}
    \vspace{-8mm}
    \caption{Amplitude of the quadrupole bar as a function of the Galactocentric radius $R$. Blue dashed curve is the most successful model reported in SBM15 and the black curve is our fitted model.}
    \label{fig:Phi2}
  \end{center}
\end{figure}

A finite size of the bar implies that the amplitude of the quadrupole bar $\Phim(R, t)$ decays at large radii as $R^{-3}$. At small radii the bar's quadrupole must vanish as fast as $R^2$ to ensure the perturbed surface density to be azimuthally smooth at the origin. Thus we model the radial dependence of the bar as 
\begin{equation}
  \Phi_m(R, t) \propto \frac{R^2}{\left[\Rb(t) + R\right]^5},
  \label{eq:bar_potential_amp_estimate}
\end{equation}
where $\Rb(t)$ is a scale length of the bar modeled to increase as the bar slows down. In concordance with \cite{athanassoula1992existence}, we model $\Rb(t)$ such as to keep the ratio against the co-rotation radius constant:
\begin{equation}
  b \equiv \frac{\Rb(t)}{\RCR(t)} = {\rm const}.
  \label{eq:Rb_RCR_ratio}
\end{equation}
The strength of the perturbation is parametrized by the ratio of the maximum azimuthal force due to the bar and the radial force due to the unperturbed potential at the co-rotation radius $\RCR$:
\begin{equation}
  A \equiv \frac{\left|\frac{1}{R}\frac{d \Phib}{d \varphi}\right|_{\RCR}}{\left|\frac{d \Phio}{d R}\right|_{\RCR}}.
  \label{eq:A}
\end{equation}
The amplitude of the bar potential then takes the following form:
\begin{equation}
  \Phi_m(R, t) = - \frac{A \vc^2}{m} \left[\frac{R}{\RCR(t)}\right]^2 \left[\frac{b + 1}{b + R/\RCR(t)}\right]^5.
  \label{eq:bar_potential_amp}
\end{equation}
The choice of negative sign ensures alignment of $\varphi = \phib$ with the bar's major axis (the potential minimum). \Figref{fig:Phi2} shows $\Phi_m$ where we fit our model to that of \citep[]{sormani2015gas3} (hereafter SBM15), which was constrained to reproduce the central Milky Way's gas flow pattern. In accordance with SBM15, we set $\Omegap = 40 \kmskpc$ corresponding to $\RCR = 5.875 \kpc$ and fit the model via $b$ and $A$. We note that our bar potential is significantly stronger near the OLR, and needs to be adapted when a quantitative fit of the resonance is required. The fitted value $b = 0.28$\footnote{This value should not be identified with the ratio between $\RCR$ and the length of the bar semimajor axis reported in \cite{athanassoula1992existence}.} is used for all simulations presented in this paper. For the strength of the bar, we run simulations with a variety of values in the range $A \in [0.01,0.03]$. In our slowing bar model, both $b$ and $A$ are kept constant while the bar slows down. 

\subsection{Review of linear perturbation theory}
\label{sec:linear_theory}

Orbits free of resonant trapping are well described by linear perturbation theory where any deviation from circular orbit is assumed to be small at the order of $\epsilon \sim \Phib/\Phio$ (see \citealt{binney2008galactic} pp.189-191 where the equation below is derived). Clearly, this assumption breaks down near resonances. Specifically, when the change of $\Omegap$ per bar rotation period is sufficiently small, we obtain the following solution for the radius of an orbit perturbed by the bar:
\begin{align}
  R(t)
  &= \Rg + \Ra \cos \left(\kappa t + \theta_{R0} \right) \nonumber \\
  &- \left[ \frac{2 \Omega \Phim}{R \left(\Omega - \Omegap\right)} + \frac{\partial \Phim}{\partial R} \right]_{\Rg} \frac{\cos m\left(\Omega - \Omegap\right)t}{\kappa^2 - m^2 \left(\Omega - \Omegap\right)^2},
  \label{eq:nonres_sol_const_R}
\end{align}
where $\Rg$ is the guiding radius and $\Omega$ is the rotation frequency of a circular orbit at $\Rg$. The second term describes the epicycle motion with amplitude $\Ra$, frequency $\kappa$ and initial phase $\theta_{R0}$. The third term oscillates with a beat frequency $m(\Omega - \Omegap)$ between the orbit and the perturbation implying that in the absence of epicycle motion, the orbit closes in the corotating frame of the bar after a beat period. These orbits with $\Ra = 0$ are termed the parent orbit of a class of orbits with identical parameters but $\Ra > 0$. Akin to a driven harmonic oscillator, the assumption of small excursion breaks down near resonances: the third term in \eqref{eq:nonres_sol_const_R} indicates a divergence of radius at the corotation resonance (CR, $\Omega = \Omegap$), the outer Lindblad resonance (OLR, $\Omega - \Omegap = - \kappa/m$), and the inner Lindblad resonance (ILR, $\Omega - \Omegap= \kappa/m$). Each divergence is accompanied by a sign change in the third term: typically the first term in the square bracket dominates, so that at each major resonance, the orientation of the parent orbits switch between alignment ($x_1$ orbits) and anti-alignment ($x_2$) with the bar major axis. Thus when the pattern speed changes, orbits passed by the major resonances switch their alignment (if not caught).

Linearizing the equation of motion has swept away the possibility of finding excitations of other high order resonances with the bar. In principle, an unlimited number of resonances occur, when commensurability is satisfied between the radial frequency $\OmegaR$ ($\kappa$ in the limit of epicycle approximation) and the azimuthal frequency $\Omegaphi$ ($\Omega$ for circular orbits) with respect to the bar pattern speed:
\begin{equation}
  \NR \OmegaR + \Nphi (\Omegaphi - \Omegap) = 0, ~~~~ \vN = (\NR, \Nphi) \in \mathbb{Z}^2.
  \label{eq:res_condition}
\end{equation}
With no loss of generality, we define $\Nphi \geq 1$ since resonance at $(\NR, -\Nphi)$ is a repetition of $(-\NR, \Nphi)$. This resonance condition depends on the bar pattern speed but not on the mode/wave number $m$ of the bar; even with a quadrupole bar ($m=2$), resonances with $\Nphi \neq m$ occur, but adding e.g. an octopole ($m=4$) affects their relative strength \citep{Monari2019signatures}. Orbits that exactly satisfy the resonant condition are closed in the co-rotating frame of the bar. Their stability -- the capability of becoming a parent orbit -- was analyzed in detail, for example, by \cite{contopoulos1989orbits}.

\subsection{Resonant dragging}
\label{sec:resonant_dragging}

We here set out to study orbits trapped and dragged by a slowing bar. Many of the results presented in this section are found in \cite{Tremaine1984Dynamical} who quantified the dynamical friction exerted on the bar by a spherical halo. Here, we focus on the behaviour of the perturbed orbits rather than their feedback on the bar.

The motion of quasi-periodic orbits is best described using actions $\vJ = (\JR, \Jphi)$, which define a torus, and corresponding angles $\vtheta = (\thetaR, \thetaphi)$, which encode the position on the surface of a torus. The main benefit of these actions is their approximate conservation under adiabatic changes; e.g. if the system slowly gains mass, or e.g. the bar slowly grows/decelerates, actions of most orbits will be conserved, with the exception of orbits with a resonant condition, or a too large orbital period violating the condition of adiabaticity. The azimuthal action $\Jphi$ is identical to the angular momentum $\Lz$, and $\JR$ is a measure of the radial excursions from a circular orbit:
\begin{equation}
  \Jphi \equiv \frac{1}{2 \pi} \oint d\varphi ~ \pphi = \Lz, ~~~ \JR \equiv \frac{1}{2 \pi} \oint dR ~ \pR.
  \label{eq:azimuthal_action}
\end{equation}
The divergence at resonance that we encountered in the linear perturbation theory can be removed by performing a canonical transformation to a frame of reference that rotates with the resonant frequency (\cite{lichtenberg1992regular} p109-117). Near but slightly off the resonance, the resonant frequency 
\begin{equation}
  \Omegas \equiv \NR \OmegaR + \Nphi (\Omegaphi - \Omegap)
  \label{eq:slow_frequency}
\end{equation}
becomes very small and thus its time integral, the so called slow angle variable,
\begin{equation}
  \thetas \equiv \NR \thetaR + \Nphi \left[\thetaphi - \int_0^t dt'~\Omegap(t') \right]
  \label{eq:slow_angle}
\end{equation}
evolves slowly around the resonance compared to $\thetaR$. The timescale disparity between $\thetas$ and $\thetaR$ enables us to separate the dynamics into slow and fast components. Thus we make a canonical transformation to a new set of angle-action variable $(\vtheta', \vJ')$ by choosing $\thetaR$ to be the other new angle which we rename as the fast angle variable:
\begin{equation}
  \theta_f \equiv \theta_R.
  \label{eq:fast_angle}
\end{equation}
To obtain the new actions $\vJ' = (\Jf, \Js)$, we perform a canonical transformation via a generating function of form $W(\vtheta, \vJ', t)$. Recall from classical mechanics that
\begin{equation}
 \vtheta' = \frac{\partial W}{\partial \vJ'}, ~~ \vJ = \frac{\partial W}{\partial \vtheta}, ~ \text{and} ~~ H'(\vtheta', \vJ', t) = H(\vtheta, \vJ, t) + \frac{\partial W}{\partial t}.
 \label{eq:canonical_transformation}
\end{equation}
The first equality instructs us how to construct the simplest $W$:
\begin{equation}
  W(\vtheta, \vJ', t) = \left\{\NR \thetaR + \Nphi \left[\thetaphi - \int_0^t dt'~\Omegap(t') \right] \right\} \Js + \thetaR \Jf.
  \label{eq:generating_function}
\end{equation}
The second equation gives
\begin{equation}
  \Jphi = \Nphi \Js, ~~~ \JR = \NR \Js + \Jf,
  \label{eq:old_action}
\end{equation}
and thus
\begin{equation}
  \Js = \frac{\Jphi}{\Nphi}, ~~~ \Jf = \JR - \frac{\NR}{\Nphi} \Jphi.
  \label{eq:new_action}
\end{equation}
The last of the three equations in (\ref{eq:canonical_transformation}) provides
\begin{equation}
  H'(\vtheta', \vJ', t)
  = H_0(\vJ') + \sum_{\vk} \Psik(\vJ', t) ~\e^{i \vk \cdot \vtheta'} - \Nphi \Omegap(t) \Js,
  \label{eq:H_new}
\end{equation}
where the perturbing potential is developed into a Fourier series $\Psik(\vJ', t)$ on the set of indices $\vk = (\kf,\ks)$ (equations for $\Psik$ are given in Appendix~\ref{app:fourier_coefficient}). During the rapid cycles in $\thetaf$, $\thetas$ can be assumed constant. Hence, one can extract the slow dynamics of $\thetas$ by averaging the Hamiltonian over $\thetaf$:
\begin{equation}
  \bH'(\thetas, \vJ', t)
  = H_0(\vJ') + \sum_{\ks \neq 0} \Psi_{\ks}(\vJ', t) ~\e^{i \ks \thetas} - \Nphi \Omegap(t) \Js,
  \label{eq:H_average}
\end{equation}
where $\Psi_{\ks} \equiv \Psi_{(0,\ks)}$. For the major resonances with $\Nphi=m$, $\Psi_{\ks}$ is non-zero only at $\ks = \pm 1$. Since the Hamiltonian must be real $(\Psi_{- \ks} = \Psi^{\ast}_{\ks})$, we have
\begin{equation}
  \bH'(\thetas, \vJ', t) = H_0(\vJ') + 2 |\Psi_1(\vJ', t)| \cos \left(\thetas + \psi_1\right) - \Nphi \Omegap(t) \Js,
  \label{eq:H_average_real}
\end{equation}
where $|\Psi_1|$ and $\psi_1$ describe the amplitude and phase of the complex Fourier coefficients $\Psi_1$. For the purpose of brevity, we henceforth use the following auxiliary variables:
\begin{equation}
  \Psi \equiv 2 |\Psi_1(\vJ', t)|, ~~~~ \theta \equiv \thetas + \psi_1.
  \label{eq:psi}
\end{equation}
The equations of motion are then
\begin{align}
  \dJs
  &= - \frac{\partial \bH'}{\partial \thetas}
  = \Psi \sin \theta,
  \label{eq:Hamilton_eq_Js} \\
  \dtheta
  &= \frac{\partial \bH'}{\partial \Js}
  = \frac{\partial H_0}{\partial \Js} + \frac{\partial \Psi}{\partial \Js} \cos\theta - \Nphi \Omegap(t).
  \label{eq:Hamilton_eq_thetas}
\end{align}
By differentiating \eqref{eq:Hamilton_eq_thetas} with respect to time, ignoring terms small to second order in $\Psi$ and also $\dthetas (= \Omegas)$ as it vanishes at the resonance, and substituting \eqref{eq:Hamilton_eq_Js}, we obtain
\begin{equation}
  \ddtheta - G \Psi \sin \theta - \frac{\partial \dot{\Psi}}{\partial \Js} \cos \theta + N_\varphi \dOmegap(t) = 0
  \label{eq:modified_pendulum}
\end{equation}
where
\begin{equation}
  G \equiv \frac{\partial^2 H_0}{\partial \Js^2}.
  \label{eq:G}
\end{equation}
We recognise equation (\ref{eq:modified_pendulum}) as a classical pendulum equation \citep[]{chirikov1979universal} with additional terms incorporating the growth of the bar (third term) and the change in pattern speed (fourth term). In our model, the order of the third term compared to the fourth term is as small as
\begin{equation}
  -\frac{\partial \dot{\Psi}}{\partial \Js} \frac{1}{N_\varphi \dOmegap} \sim \frac{-\dot{\Psi}}{\Jphi \dOmegap} \sim \frac{-A \vc^2}{m} \frac{\vCR}{R} \frac{1}{R \vc \dOmegap} = \frac{A}{m},
  \label{eq:ordering_elongation_term}
\end{equation}
so as first order approximation we will neglect the third term. The third term will become non-negligible when the strength of the bar is modeled to grow rapidly (in our current model we assumed $A = {\rm const}$, so $\dot{\Psi}$ is due only to the stretching of the bar). We leave exploration of a slowing + strengthening bar to a later study.

\begin{figure*}
  \begin{center}
    \setlength\columnsep{0pt}
    \begin{multicols}{3}
      \includegraphics[width=5.8cm]{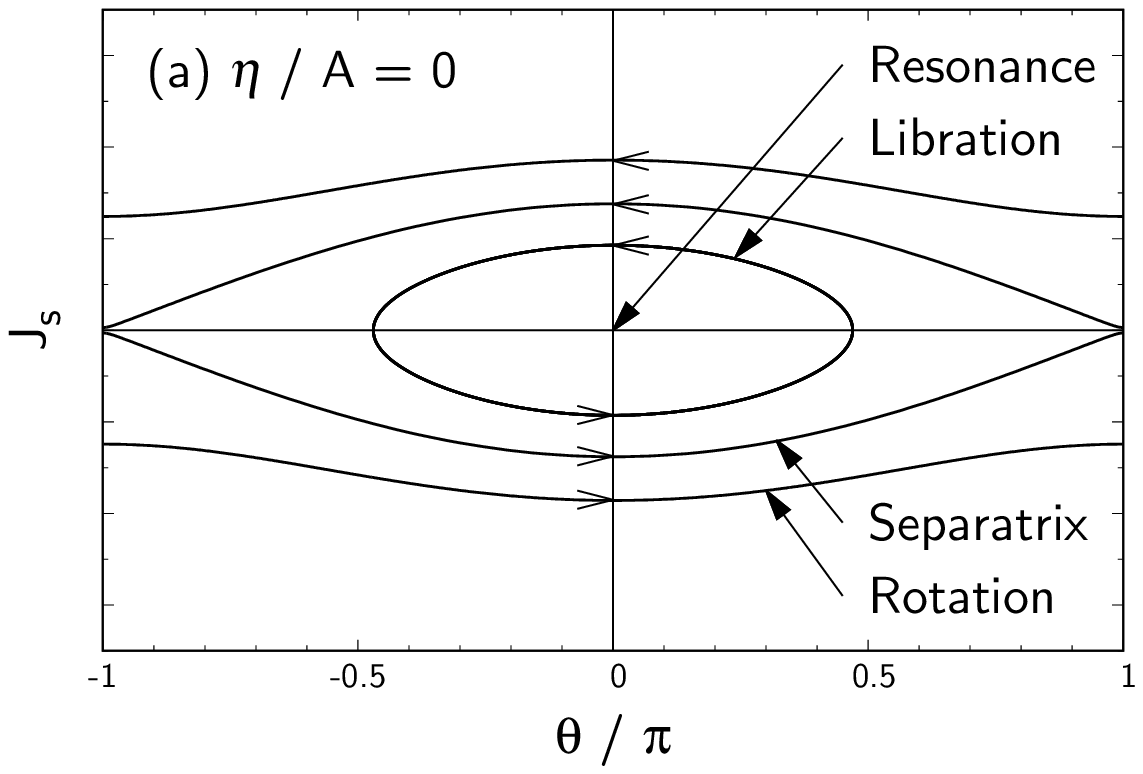}
      \newpage
      \includegraphics[width=5.8cm]{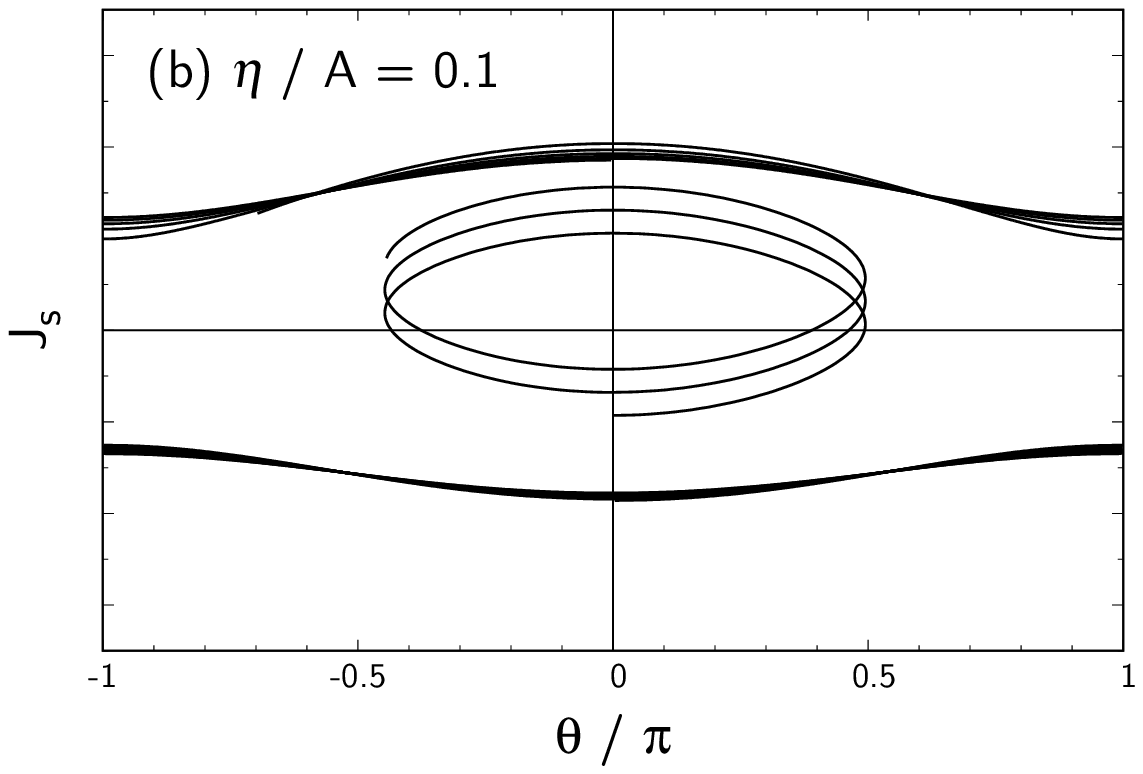}
      \newpage
      \includegraphics[width=5.8cm]{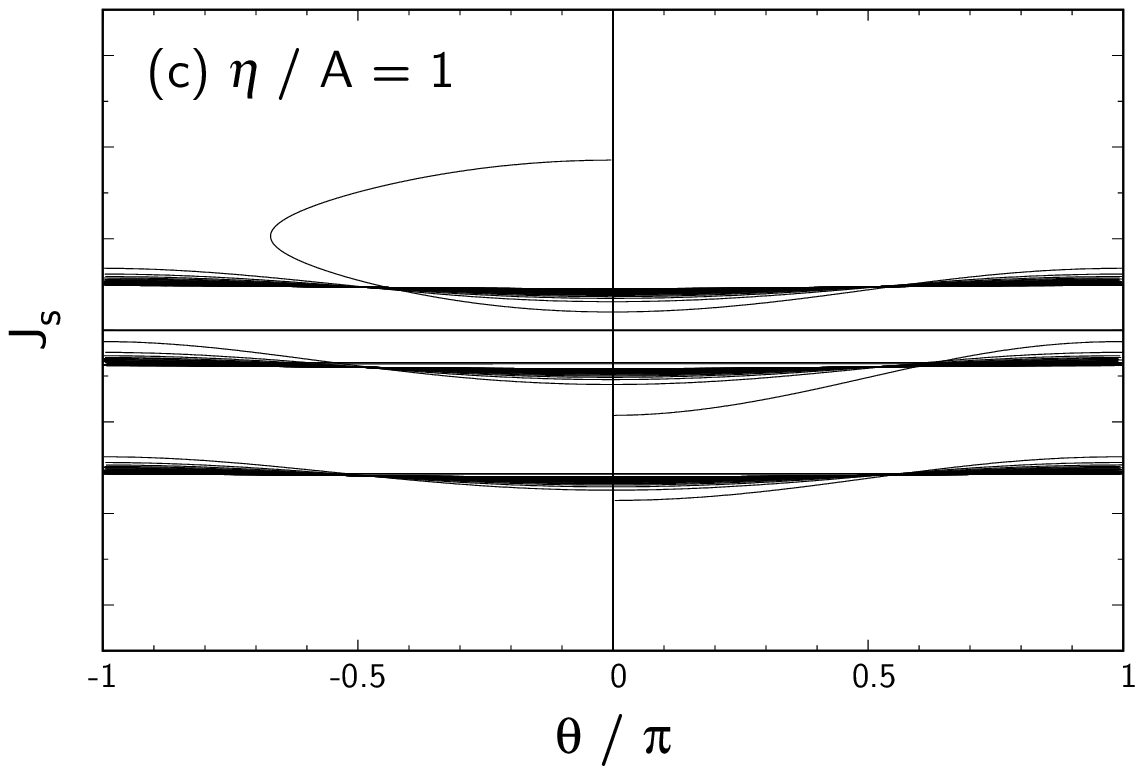}
    \end{multicols}
    \setlength\columnsep{20pt}
    \vspace{-6mm}
    \caption{Understanding the motion near resonance in terms of pendulum dynamics. The plots are drawn by numerically integrating \eqref{eq:modified_pendulum_simplified}. (a) In a constantly rotating bar ($\eta/A = 0$), $\theta$ of trapped orbits librates around the resonance while that of non-trapped orbits circulates above or below the separatrix \citep[]{lichtenberg1992regular}. (b) When the bar slows down slowly such that $0 < \eta/A < 1$, trapped orbits can be resonantly dragged towards higher $\Js$. (c) Orbits cannot stay in resonance when the resonance sweeps too fast ($\eta/A \geq 1$).}
    \label{fig:pendulum}
  \end{center}
\end{figure*}

We now look at the impact of the dragging/slowing term on the modified pendulum equation (\ref{eq:modified_pendulum}). Using $\Psi$ and $G$ from Appendix~\ref{app:fourier_coefficient} and \ref{app:calculate_G}, the order of the slowing term $N_\varphi \dOmegap(t)$ is
\begin{equation}
  \frac{\Nphi \dOmegap}{G \Psi} \sim \frac{\Nphi \dOmegap}{\left(- \frac{\Nphi^2}{\RCR^2} \right) \left(\delta_{m \Nphi}\frac{A \vc^2}{m}\right)} = \frac{\eta}{A}
  \label{eq:ordering}
\end{equation}
where we assumed $G < 0$ and invoked the parameter $\eta = - \dOmegap/\Omegap^2$ defined in \eqref{eq:slowing_rate}. In the limit $\JR \rightarrow 0$, the approximation made above is exact at the CR but underestimates by a factor of $\sim0.83$ at the OLR. Using \eqref{eq:ordering}, we may rewrite \eqref{eq:modified_pendulum} as
\begin{equation}
  \ddot{\theta} - G \Psi \left(\sin \theta - \frac{\eta}{A} \right) = 0.
  \label{eq:modified_pendulum_simplified}
\end{equation}
Note that the sign of the slowing term $\eta/A$ reverses at the ILR where $G$ becomes positive. In the following, we approximate $G$ and $\Psi$ with their values at the resonance $\Js = \Jsres$ at the time of capture $t = \tres$ on the assumption that their time evolution is slow compared to that of $\theta$. The corresponding $\Jf$ is determined by the resonance condition $\Omegas(\Jsres, \Jf) = 0$. We numerically integrate \eqref{eq:modified_pendulum_simplified} together with (\ref{eq:Hamilton_eq_Js}) and follow the motion of orbit in the $(\theta, \Js)$ plane. \Figref{fig:pendulum} (a) shows the phase plane of a pendulum with $\eta/A = 0$. As described in the figure, trapped orbits librate around the resonance at $(\theta, \Js) = (0, \Jsres)$. This region is bounded by the separatrix, which has maximal/infinite libration period. Outside the separatrix, non-trapped orbits freely circulate, with less amplitude in $\Js$ the further they are from the resonance. Figures \ref{fig:pendulum} (b) and (c) show the same plot when the bar slows down moderately ($0 < \eta/A \ll 1$) and extremely rapidly ($\eta/A = 1$). As in Fig. \ref{fig:pendulum} (a) the amplitude of oscillations in non-trapped orbits depends on the proximity to the resonant region, so fluctuations of orbits circulating above the separatrix amplifies as the resonance approach while that below the separatrix attenuates as the resonance pass away. On the other hand, in the librating regime, the additional term $\eta/A$ causes a drift in $\Js$. To see how this works, let us employ the small-angle approximation. We then obtain
\begin{equation}
  \theta = \hat\theta \cos \left(\omega t + \phi \right) + {\rm sgn}(-G) \frac{\eta}{A}, ~~ {\rm where} ~~ \omega \equiv \sqrt{|G| \Psi}.
  \label{eq:pendulum_approx_sol}
\end{equation}
$\hat{\theta}$ and $\phi$ are the amplitude and initial phase of the oscillation. To justify the small-angle approximation, we require $\eta/A$ to be small which is satisfied when the bar is either strong or slowing down slowly. We insert this solution into \eqref{eq:Hamilton_eq_Js} and integrate:
\begin{align}
  \Js
  &= \Psi \Biggl\{ \cos \left[ {\rm sgn}(-G) \frac{\eta}{A} \right] \int dt~ \sin \left[ \hat\theta \cos \left( \omega t + \phi \right) \right] \nonumber \\
  &\hspace{0.5cm} + \sin \left[ {\rm sgn}(-G) \frac{\eta}{A} \right] \int dt~ \cos \left[ \hat\theta \cos \left( \omega t + \phi \right) \right] \Biggr\}.
  \label{eq:J_s_dot}
\end{align}
Clearly, this describes an oscillation (first term) plus a small drift (second term) of the orbit in $\Js$ along with the resonance. When $\eta = 0~(\dOmegap = 0)$, the drift term vanishes. When $\eta > 0~(\dOmegap < 0)$ we find that, at the OLR and the CR $(G < 0)$, it leads to a positive drift in $\Js$ and thus in $\Jphi$; trapped orbits at the OLR and the CR are dragged radially outwards by the slowing bar. In contrast, at the ILR $(G > 0)$, resonant orbits are dragged towards lower $\Jphi$. \Figref{fig:pendulum} (c) shows that if the bar is too weak and/or the resonant sweeping is too fast, the third term of \eqref{eq:modified_pendulum_simplified} dominates the dynamics and will force $\theta$ to circulate. In such case, orbits cannot stay trapped at resonance and thus dragging will not occur.

\begin{table}
	\centering
	\caption{Summary of the direction of resonant dragging due to decrease in bar pattern speed. The sign of $\dJphi$ is determined by the sign of the non-linearity parameter $G$, and $\dJR$ is related to $\dJphi$ by \eqref{eq:\JR_evolution}.}
	\label{tab:dJdt_summary}
	\begin{tabular}{lccc}
		\toprule
		& ILR & CR & OLR\\
		\hline
		$\dot{J}_\varphi$ & - & + & +\\
		$\dot{J}_R$ & + & 0 & +\\
		\bottomrule
	\end{tabular}
\end{table}

On averaging the Hamiltonian over the fast angle, we have implicitly concluded that the fast action is effectively conserved ($\dJf = 0$). Therefore any change in angular momentum will be accompanied by a change in the radial action:
\begin{equation}
  \dJR = \frac{\NR}{\Nphi} \dJphi.
  \label{eq:\JR_evolution}
\end{equation}
This is still the well-known result known to most readers in the context of radial migration \citep[]{Sellwood2002radial}: as a response to the positive dragging in $\Jphi$, the radial action of trapped orbits is conserved at CR ($\NR / \Nphi = 0$) whereas increases at OLR ($\NR / \Nphi > 0$). On the other hand, a negative dragging in $\Jphi$ at ILR ($\NR / \Nphi < 0$) will be compensated by an increase in radial action. This explains why \cite{weinberg1994kinematic} observed a large increase in velocity dispersion near the OLR. We summarise the direction of resonant dragging in table \ref{tab:dJdt_summary}. These behaviours are confirmed numerically in \Figref{fig:slow_orbit_Lz_JR}.

We can also understand the effect of the slowing term from the viewpoint of an effective potential. By multiplying $\dtheta$ on the modified pendulum equation (\ref{eq:modified_pendulum_simplified}) and integrating over time, we obtain the following energy integral:
\begin{align}
  \Ep \equiv \frac{1}{2}\dtheta^2 + V(\theta), ~~ {\rm where} ~~ V(\theta) \equiv \omega^2 \left(- \cos \theta - \frac{\eta}{A} \theta \right).
  \label{eq:pendulum_energy}
\end{align}
Note that the pendulum energy $\Ep$ is not conserved under slow changes in $\omega$. The true adiabatic invariant is the action of libration
\begin{equation}
  \Jl \equiv \frac{1}{2\pi} \oint d\theta \Js(\theta),
  \label{eq:libration_action}
\end{equation}
which quantifies the amplitude of motion in the slow angle-action plane. It is approximately conserved if the libration period $\Tl$ is sufficiently shorter than the migration timescale of the resonance $\Tres$. Apart from the vicinity of the separatrix, $\Tl$ is of order
\begin{align}
  \Tl \sim \frac{2\pi}{\omega} \sim \frac{1}{\sqrt{A}\Omegap},
  \label{eq:libration_period}
\end{align}
while the time for a resonance to move by its width (eq.~\ref{eq:DJmax}) is
\begin{align}
  \Tres = \frac{\DJsmax}{\dJs} \sim \frac{\sqrt{A}}{\eta\Omegap}.
  \label{eq:libration_period}
\end{align}
Thus $\Jl$ is conserved for most orbits when $\Tl/\Tres \sim \eta/A$ is small.

\begin{figure}
  \begin{center}
    \includegraphics[width=8.5cm]{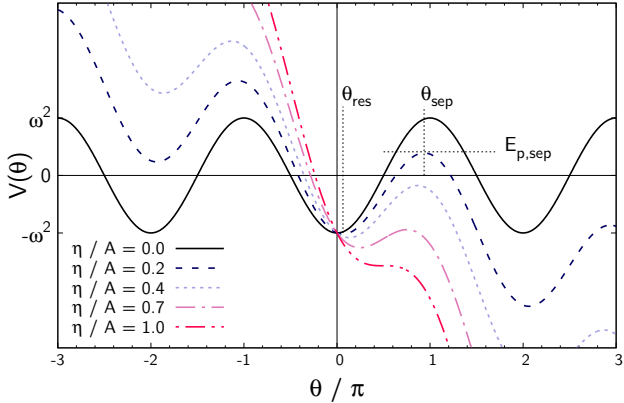}
    \vspace{-6mm}
    \caption{Potential of the modified pendulum. Orbits trapped by the resonance are confined within the potential well. A decreasing pattern speed results in a tilt in the potential and thus the minimum $\Ep$ necessary to escape the resonance becomes smaller than that with constant $\Omegap$.}
    \label{fig:pendulum_V}
  \end{center}
\end{figure}

Figure~\ref{fig:pendulum_V} shows the configuration of the effective potential $V(\theta)$ which is inclined due to the $\eta/A$ term in \eqref{eq:pendulum_energy}. The resonance centre $\thetares$ and the angle of the local maximum $\thetasep$ reached by stars orbiting along the separatrix are: 
\begin{align}
  \thetares &= \sin^{-1}\left(\frac{\eta}{A}\right), ~~~ 0 \leq \thetares \leq \frac{\pi}{2},
  \label{eq:thetares} \\
  \thetasep &= \sin^{-1}\left(\frac{\eta}{A}\right), ~~~ \frac{\pi}{2} \leq \thetasep \leq \pi.
  \label{eq:thetasep}
\end{align}
This positive shift of $\thetares$ is barely noticeable in Fig.~\ref{fig:pendulum} (b), but prominently tilts the equilibrium angle in Fig.~\ref{fig:slow_orbit_Lz_JR}. Further, the maximum $\Ep$ and $\Jl$ at the separatrix are reduced by the $\eta/A$ term:
\begin{align}
  \Epsep &\equiv \omega^2 \left(- \cos\thetasep - \frac{\eta}{A}\thetasep \right),
  \label{eq:Ep_sep} \\
  \Jlsep &\equiv \frac{2 \omega}{\pi |G|} \int_C d\theta \sqrt{2\left[\cos\theta - \cos\thetasep + \frac{\eta}{A} \left(\theta - \thetasep \right)\right]}.
  \label{eq:Jl_sep}
\end{align}
If $\Ep < \Epsep$ orbits are trapped and dragged, otherwise they will escape the resonance from $\thetasep$ and enter the lower circulating regime. The potential barrier of trapped orbits decreases as the bar slows down more rapidly. Beyond the critical value $\eta/A = 1$ where $\thetares=\thetasep=\pi/2$, the potential does not form a local extremum and thus orbits can no longer stay trapped in resonance.

\begin{figure}
  \begin{center}
    \includegraphics[width=8.5cm]{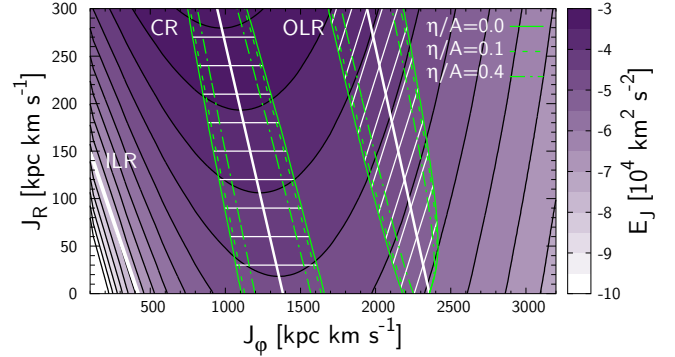}
    \vspace{-6mm}
    \caption{Thick white lines mark the location where the resonance condition is exactly satisfied and green lines indicate the maximum excursion of trapped orbits for different $\eta/A$. The color map shows the Jacobi integral $\EJ$. At the resonances, the lines of constant $\Jf$ (thin white lines) are tangent to the contours of $\EJ$. The parameters used are $A=0.02$ and $\Omegap = 40 \kmskpc$.}
    \label{fig:J_HJ}
  \end{center}
\end{figure}

A decrease in $\Epsep$ implies that the phase space volume of resonance shrinks when the bar slows down. \Figref{fig:J_HJ} shows this in action space (using \ref{eq:app_actions}-\ref{eq:app_apsis}). The thick white lines are the resonance lines, and the solid green lines mark the maximum excursion of trapped orbits from the exact resonance $\DJsmax \equiv (\Js - \Jsres)_{\rm max}$, which happens at $\Ep = \Epsep$ and $\theta = \thetares$:
\begin{equation}
  \DJsmax = \frac{\omega}{|G|} \sqrt{2 \left[\cos\thetares - \cos\thetasep + \frac{\eta}{A}\left(\thetares - \thetasep \right)\right]}.
  \label{eq:DJmax}
\end{equation}
The resonant volumes shrink with increasing $\eta/A$, and vanish when $\eta/A$ reaches unity. Note that not all orbits within the green boundary are trapped, as the trapping also depends on the angles $\vtheta$.

The white thin lines crossing the resonance represent the line of constant $\Jf$ which the librating orbits are assumed to follow in our resonance theory. In fact, conservation of $\Jf$ is truly satisfied only at the resonance line ($\vJ = \vJres$) and is otherwise an approximation to the precise conservation of the Jacobi integral $\EJ$ which is mapped by the colour scale in \Figref{fig:J_HJ} as in \cite{binney2017orbital}, where he computed the actions in a 3D axisymmetric potential using torus mapping. The lines of constant $\Jf$ and $\EJ$ match precisely at the resonance line but deviate for large libration amplitude which is most notable at the CR. Mathematically, conservation of $\EJ = E - \Omegap \Jphi$ requires
\begin{align}
  \Delta \EJ
  &= \frac{\partial E}{\partial \JR} \Delta \JR + \frac{\partial E}{\partial \Jphi} \Delta \Jphi  - \Omegap \Delta \Jphi \nonumber \\
  &= \OmegaR \Delta \JR + (\Omegaphi  - \Omegap) \Delta \Jphi = 0
  \label{eq:HJ_const}
\end{align}
which becomes equivalent to $\Delta \Jf = \Delta \JR - \NR / \Nphi \Delta \Jphi = 0$ only when the frequencies $\vOmega(\vJ)$ are approximated by their values at $\vJres$.

\section{Test-particle simulation}
\label{sec:test_particle_sim}

To ensure full control over the model parameters, we use a test-particle simulation in an analytical potential presented in Section \ref{sec:model}. Our simulation technique is similar to \cite{muhlbauer2003kinematic}, who examined the kinematics around a steadily rotating bar. We integrate in each simulation $10^8$ particles forward in time using a 4th order symplectic integrator \citep[]{yoshida1993recent}, with a time step of $0.1 \Myr$. Parameters of our model are summarised in table \ref{tab:summary_variable}.

\subsection{Initial distribution function}
\label{sec:initial_DF}

The initial distribution function is given by \citep[]{dehnen1999simple}:
\begin{equation}
  f(E,\Lz) \propto \frac{\Sigma(R_E)}{\sigma_R^2(R_E)} \exp\left[ - \frac{\Omega(R_E)\left[\Lc(E) - \Lz\right]}{\sigma_R^2(R_E)} \right],
  \label{eq:Dehnen_DF}
\end{equation}
where $R_E$, $\Omega(R_E)$, $\Lc(E)$ is the radius, circular frequency, and angular momentum of a circular orbit with energy $E$. We assume an exponential profile with scale lengths $R_{\Sigma}$ and $R_{\sigma}$ for both the surface density $\Sigma(R)$ and the radial velocity dispersion $\sigma_R(R)$ of the disk:
\begin{equation}
  \Sigma(R) = \Sigma_\odot \e^{- (R - R_0) / R_\Sigma}, \quad \sigma_R(R) = \sigma_\odot \e^{- (R - R_0) / R_\sigma},
  \label{eq:velocity_dispersion}
\end{equation}
where $R_0$ is the galactocentric distance of the Sun and $\sigma_\odot$ is the local stellar velocity dispersion. Throughout our work, we adopt $R_\Sigma$ = 2.5 kpc, $R_\sigma$ = $R_0$ = 8.2 kpc, and $\sigma_\odot \equiv \sigma_R (R_0) = 40 \kms$. 

Once $(E, \Lz)$ are determined from \eqref{eq:Dehnen_DF}, we compute the initial radius $R$ by integrating \eqref{eq:app_angles} up to a random value of $\thetaR \in [0,2\pi)$. The corresponding initial velocities are then determined $\vphi = \Lz/R ~, \vR^2=2[E-\Phi(R)]-\Lz^2/R^2$ and the initial azimuthal angle is sampled randomly from $\varphi \in [0,2\pi)$.

\subsection{Adiabatic growth of the bar}
\label{sec:adiabatic_growth}

A sudden onset of the bar will permanently change the actions. As in the past literature, we avoid such an unnecessary distortion from a more realistic case by growing the bar slowly, i.e. we ramp up its strength $A$ from $0$ to its final value $\Af$ during the time interval $0 < t < t_1$ using the polynomial law from \citep[]{dehnen2000effect}:
\begin{equation}
  A(t) = \Af \left( \frac{3}{16}\xi^5 - \frac{5}{8}\xi^3 + \frac{15}{16}\xi + \frac{1}{2} \right), ~~~ \xi = \frac{2t}{t_1} - 1.
  \label{eq:adiabatic_growth}
\end{equation}
Choosing $t_1 = 2 \Gyr$, this ramp is adiabatic for most orbits, apart from those near the surface/separatrix of the resonance where the libration period diverges.

\subsection{Pattern speed}
\label{sec:pattern_speed}

\begin{figure}
  \begin{center}
    \includegraphics[width=8.5cm]{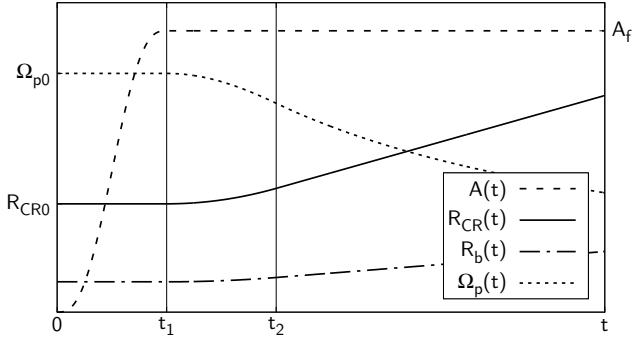}
    \vspace{-6mm}
    \caption{Schematic diagram of the strength of the bar $A(t)$, the co-rotation radius $\RCR(t)$, the bar length $\Rb(t)$, and the pattern speed $\Omegap = \vc / \RCR(t)$. The bar is adiabatically grown while keeping the pattern speed constant, and subsequently slowed down with its strength unchanged. The length of the bar is elongated proportional to the co-rotation radius as $\Rb(t) = b \RCR(t)$ where $b = 0.28$ is determined by fitting our model to SBM15.}
    \label{fig:A_RCR_wb}
  \end{center}
\end{figure}

As described in section \ref{sec:model}, we model the pattern speed to decrease inverse proportional with time which amounts to a linear increase in co-rotation radius $\RCR$. To separate effects, we keep $\RCR$ constant during the ramp-up of the bar amplitude ($0 < t < t_1$), and then smoothly start the slowing within ($t_1 < t < t_2$):
\begin{equation}
  \RCR(t) =
  \begin{cases}
    & \RCRo \hspace{3.75cm} (0 < t < t_1) \\
    & \RCRo + \frac{1}{2} \vCR \frac{(t - t_1)^2}{t_2 - t_1} \hspace{1.8cm} (t_1 < t < t_2) \\
    & \RCRo + \frac{1}{2} \vCR (t_2 - t_1) + \vCR (t - t_2) \hspace{0.5cm} (t_2 < t)
  \end{cases}
  \label{eq:corotation_radius_model}
\end{equation}
where $\RCRo$ is the initial co-rotation radius and $\vCR$ is the velocity of the co-rotation radius (here typically of the order of $0.1 - 1 \kms$). Remind that in line with the decrease in pattern speed, the bar is made more elongated by keeping the linear relation $\Rb = b \RCR$. The time variation of the bar's properties are drawn in \Figref{fig:A_RCR_wb}.

\subsection{Selection function}
\label{selection_function}

\begin{figure}
  \begin{center}
    \includegraphics[width=8cm]{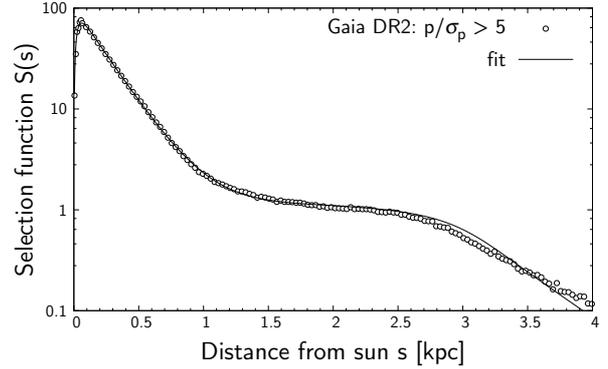}
    \vspace{-2mm}
    \caption{Distance selection function of \textit{Gaia} DR2 fitted with \eqref{eq:selection_function}. We apply this to our simulation to assess the impact of the selection effects.}
    \label{fig:selection_function_log}
  \end{center}
\end{figure}

When we compare our model with observational data, we apply to our simulation the distance-dependent selection function of the \textit{Gaia} data with quality cut in parallax of $p/\sigpar > 5$ \citep[]{Schoenrich2019distance}. The adopted selection function is shown in \Figref{fig:selection_function_log}. The data are fitted using the following analytical function $S(s)$:
\begin{align}
  S(s) &= a_0 A(s) B(s) C(s),
  \label{eq:selection_function} \\
  A(s) &= \exp(- a_1 s) + \frac{a_2 \exp(- a_3 s)}{1 + \exp\left[- a_4 (s - a_5)\right]}, \nonumber \\
  B(s) &= \frac{\tan^{-1}\left[a_7 (a_8 - s) + a_6 \right]}{\pi/2 + a_6}, ~~ 
  C(s) = 1 - \exp(- a_9 s), \nonumber
\end{align}
where $s$ is the distance from the Sun and $a_i$ ($i$=0...9) are fitting parameters. Two things are to be noted here: 
\begin{itemize}
 \item As a somewhat trivial point, the function here is similar but not identical to the function provided in equation (6) of \cite{Schoenrich2019distance}, as here we have to figure in the additional effect of the parallax cut, which must not be applied to the distance estimation.
 \item More importantly, this is only an indicative bias. In truth, the sample selection is based on a photometric selection, which will result in strong biases along age and metallicity, which are much too complex for coverage in this exploratory study. These effects will also be distance dependent, as the near field ($s < 1 \kpc$) is dominated by dwarf/subgiant stars, which have a very different age-metallicity selection function from the giant branches dominating the far field.
\end{itemize}

\begin{table}
	\centering
	\caption{Summary of parameters used in our slowing bar model. Parameters to be varied are $A$, $\eta$, and $\Omegapo$.}
	\label{tab:summary_variable}
	\begin{tabular}{lccc}
		\toprule
		Parameter & Symbol & Value\\
    \hline
    \multicolumn{3}{c}{Parameters for the slowing bar} \\
		\hline
		Bar wave number & $m$ & 2 \\
		Bar angle w.r.t. the Sun & $\phib - \phisun$ & $30^\circ$ \\
		Bar strength & $A$ & 0.01 - 0.03 \\
		Bar scale length ratio & $b \equiv \Rb/\RCR$ & 0.28 \\
		Bar slowing rate & $\eta = - \dOmegap/\Omegap^2$ & 0.001 - 0.0055 \\
		Bar initial pattern speed & $\Omegapo$ & 60 - 100 $\kmskpc$ \\
		Bar growth time & $t_1$ & 2 $\Gyr$ \\
		Transition time from constant & $t_2 - t_1$ & 1 $\Gyr$ \\
		to linear increase in $\RCR$ & &  \\
		\hline
    \multicolumn{3}{c}{Parameters for the Galactic disk} \\
		\hline
		Circular velocity & $\vc$ & $235 \kms$ \\
		Disk scale length & $R_\Sigma$ & $ 2.5 \kpc$ \\
		Local velocity dispersion & $\sigma_R(R_0)$ & $ 40 \kms$ \\
		$\sigma_R$ scale length & $R_\sigma$ & $R_0$ \\
		\bottomrule
	\end{tabular}
\end{table}

\section{Results and Discussions}
\label{sec:results}

\subsection{Constantly rotating bar}
\label{sec:constantly_rotating_bar}

Before we turn to the main topic of our paper, i.e. the effects of bar deceleration, we discuss the simpler case of a constantly rotating bar. Here we choose amplitude $A$ = 0.02 and pattern speed $\Omegap = 40 \kmskpc$ according to SBM15.

\begin{figure}
  \begin{center}
    \includegraphics[width=7.7cm]{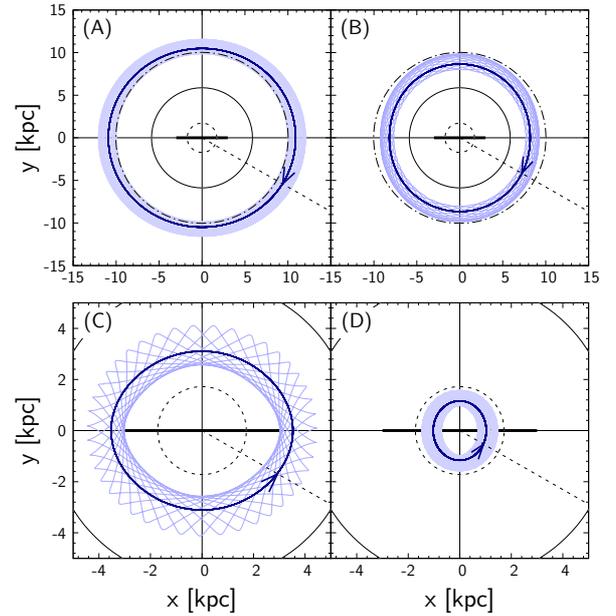}
    (a)~Non-resonant orbits.
    \includegraphics[width=7.7cm]{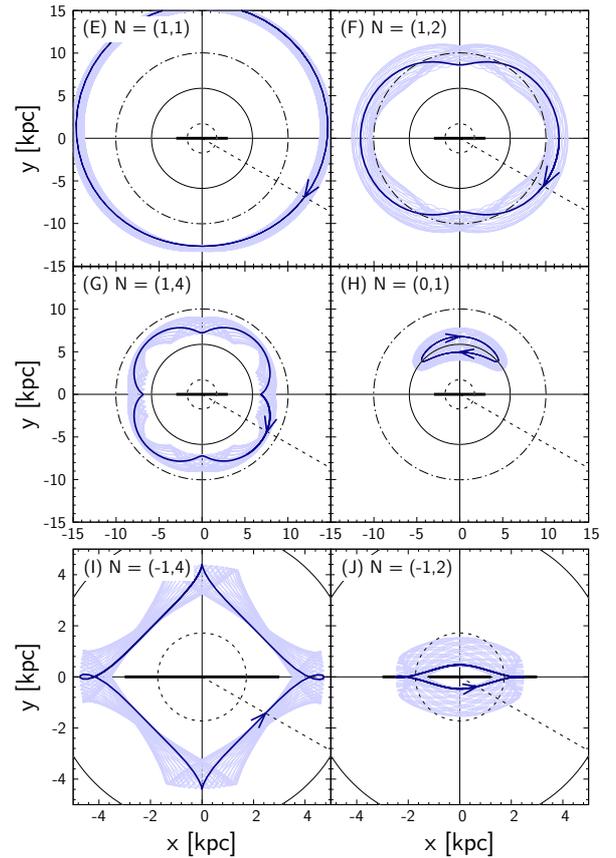}
    (b)~Resonantly trapped orbits.
    \caption{Typical orbits in a co-moving frame of a $m$ = 2 bar rotating with a steady pattern speed $\Omegap = 40 \kmskpc$. Upper figure shows orbits free from resonance and lower figure shows orbits trapped at some of the major resonances. Non-closed orbits (light blue) are parented by stable closed orbits (dark blue). The black dotted, solid, and dot-dashed circles are the ILR, CR, and OLR radii respectively. The black horizontal line is the bar's long axis and the black dotted line indicates the Solar azimuth at $\phib - \phisun = 30^\circ$.}
    \label{fig:orbitRF_a10_wb40}
  \end{center}
\end{figure}

\Figref{fig:orbitRF_a10_wb40} shows examples of different classes of orbits seen in the frame corotating with the bar, where the bar's major axis is represented by a thick black line along the $x$ axis. Note the different scale of each panel. The black circles mark the radii of ILR (dotted), CR (solid) and OLR (dot-dashed). We show non-resonant orbits in the top panel. The non-closed orbits (light blue) with non-zero $\JR$ surround their parent/closed orbits with the same $\Lz$ (dark blue). As discussed in Section \ref{sec:linear_theory}, the orbit orientation changes at each major resonances: orbits are elongated parallel to the bar (A) outside OLR, and (C) between CR and ILR, while they are elongated perpendicular to the bar (B) between CR and OLR, and (D) inside ILR. The bottom panel of Fig. \ref{fig:orbitRF_a10_wb40} shows examples of orbits trapped at the main resonances: (E) the outer 1:1 resonance, (F) the outer Lindblad resonance, (G) the outer ultra-harmonic resonance, (H) the corotation resonance, (I) the inner ultra-harmonic resonance, (J) and the inner Lindblad resonance. The corresponding resonant closed parent orbits, again depicted in dark blue, are far from circular and are beyond the description of \eqref{eq:nonres_sol_const_R}. 

\begin{figure}
  \begin{center}
    \setlength\abovecaptionskip{4pt}
    \includegraphics[width=8.2cm]{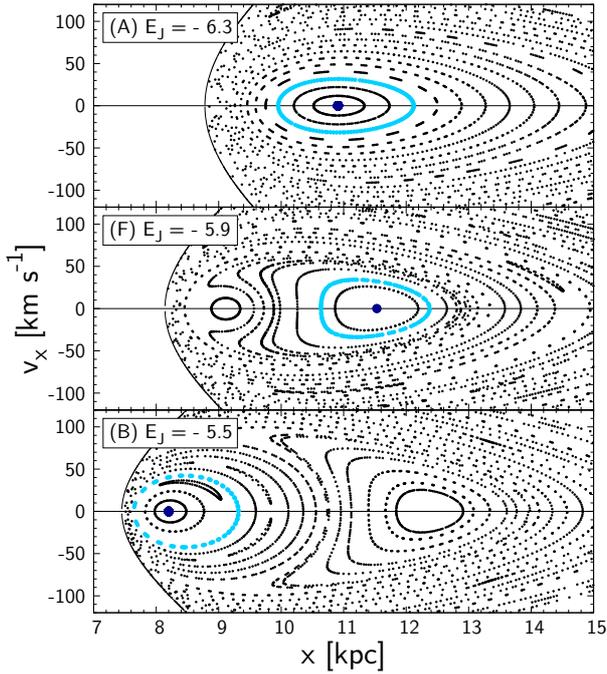}
    \vspace{-2mm}
    \caption{Surface of sections near OLR for $\EJ$ = -6.3, -5.9, and -5.5 ${\rm km}^2{\rm s}^{-2}$ respectively. The colored invariant curves correspond to orbits presented in \Figref{fig:orbitRF_a10_wb40} (A), (F), and (B). Non-closed orbits (light blue) encircle their corresponding closed parent orbits (dark blue). The phase space is restricted to the right of the thick limiting curve defined by the equation $(v_y - \Omegap x)^2 = -v_x^2 + \Omegap^2 x^2 + 2[\EJ - \Phi(x)] = 0$.}
    \label{fig:sos_const_bar}
  \end{center}
\end{figure}

The orbits' family relations at resonance are better understood using their surfaces of section. In Fig. \ref{fig:sos_const_bar}, we show surfaces of section in the reduced phase-space $(x, v_x)$ at $y=0$ and $v_y < 0$ near the OLR. Each panel shows a set of orbits with the same Jacobi integral, $\EJ = E - \Omegap \Lz$, indicated in the top left corner. Each non-closed orbit forms consequents, which appear as ring-like features in these plots (though subsequent passages are not adjacent to each other). Each colored invariant corresponds to an orbit shown in \Figref{fig:orbitRF_a10_wb40} (A), (B), and (F). Near the OLR, increasing $\EJ$ from top to bottom generally moves the mapped phase space towards lower $\Lz$ and higher $\JR$ as can be recognised in \Figref{fig:J_HJ}. The orbits in the top plot have too small $\EJ$ to reach the OLR, so they all belong to the same $x_1$ non-resonant parent orbit. The larger $\EJ$ in the middle panel allows the contour of $\EJ$ in action space (\Figref{fig:J_HJ}) to cross the OLR line. Consequently, we now see three different types of orbits: non-resonant $x_2$ orbits below the lower separatrix with small $\JR$ circulating near $x \sim 9.2 \kpc$, resonant orbits inside the separatrix around $x \sim 11.2 \kpc$, and non-resonant $x_1$ orbits above the upper separatrix with large $\JR$ surrounding the other two groups. In the bottom panel the region of $x_2$ orbits around $x \sim 8.3 \kpc$ has expanded, and the resonant domain at $x \sim 12.5 \kpc$ starts shrinking. Also, there are minor resonances occupying much smaller regions of phase space, e.g. the small crescent shape belonging to the 2:3 resonance.

\begin{figure}
  \begin{center}
    \includegraphics[width=8.0cm]{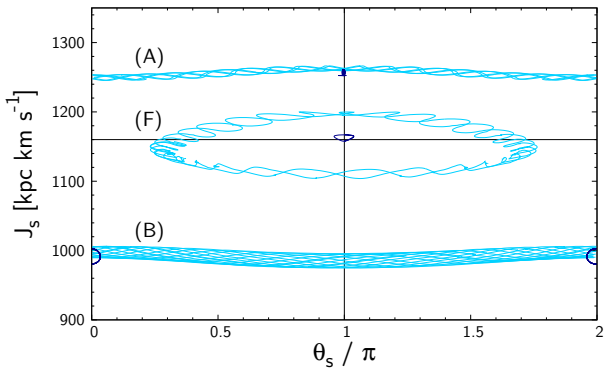}
    \vspace{-3mm}
    \caption{Motion of orbits near the OLR in the slow angle-action plane. The three curves (A), (F), and (B) correspond to orbits beyond, trapped at, and below the OLR in correspondence with Figs.~\ref{fig:orbitRF_a10_wb40} and \ref{fig:sos_const_bar}. Resonantly trapped orbits librate about the stable equilibrium point while orbits free from resonances circulate above and below the separatrix.}
    \label{fig:slowAA_const_bar}
  \end{center}
\end{figure}

Resonant and non-resonant orbits fundamentally differ in their apsidal motion \citep[]{weinberg1994kinematic,monari2017distribution}, which closely relates to the slow angle variable $\thetas$; if one defines $\thetaR = 0$ at the pericentre and writes $\varphi_\mathrm{peri} = \thetaphi - \phib$, then $\thetas = \NR \thetaR + \Nphi (\thetaphi - \phib) = \Nphi \varphi_\mathrm{peri}$. As an example, \Figref{fig:slowAA_const_bar} depicts orbits near the OLR in slow angle-action plane. At the OLR, the azimuth of the pericentre oscillates around the bar minor axis ($\varphi = \pi/2$) so the equilibrium point of $\thetas$ is $\pi$. The azimuth of the apsis of non-resonant orbits (A and B) in the bar rotating frame circulates, while that of resonantly trapped orbit (F) oscillates in a finite range. The small oscillations on top of the slow dynamics are the fast oscillations over which we have averaged in our resonant theory. As an artifact of the parent orbit elongation, the dark blue parent orbits of the non-resonant orbits are not circling in $\thetas$. The problem here is that non-resonant closed orbits in a barred potential perform two radial oscillations per rotation around the bar, so their $\thetaR$ linearly increases at rate $\pm 2(\Omegaphi - \Omegap)$ resulting in $\Omegas = \dthetas = 0$ despite them being free of resonance. In contrast, the true resonances at OLR/ILR occur when the radial oscillations on top of the closed parent orbits have frequencies equal to $\pm 2(\Omegaphi - \Omegap)$. The artifact disappears when the amplitude of radial oscillation with respect to the closed parent orbit becomes larger than the radial distortion of the closed parent orbit itself. This problem arises from the mapping to angle-action variables of an unperturbed potential. The truly conserved radial action would quantify the extent of radial distortion from the closed elongated orbit rather than from the circular orbit and the linearly increasing radial angle variable would measure the phase of radial motion relative to the parent orbits.

\begin{figure}
  \begin{center}
    \includegraphics[width=8.8cm]{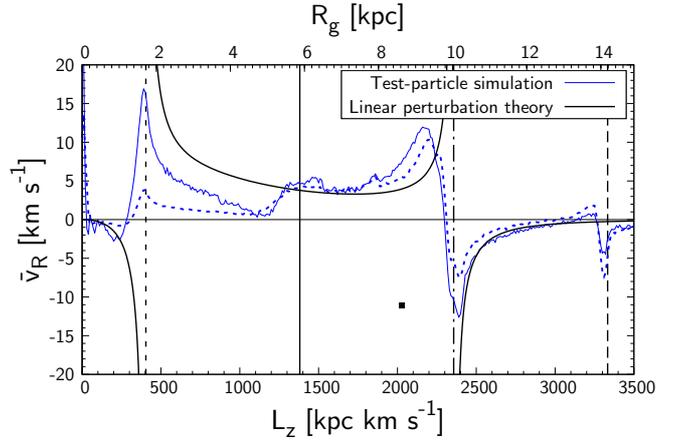}
    \vspace{-7mm}
    \caption{Mean radial velocity $\ovR$ vs. $\Lz$ at Solar azimuth. Linear perturbation theory (solid black curve, eq.\ref{eq:mean_vR}) is compared to test-particle simulations without (solid blue) and with (dashed blue) the \textit{Gaia} selection function applied. Black vertical lines represent, from left to right, ILR, CR, OLR, and O11R respectively. Black square marks the coordinates of the Sun. Since the majority of orbits are non-resonant and near circular, linear theory qualitatively captures the main features of the simulations.}
    \label{fig:LzRvR_a10_wb40_s40}
  \end{center}
\end{figure}

Now we look into the velocity distribution of an ensemble of these test-particles. \Figref{fig:LzRvR_a10_wb40_s40} shows the mean radial velocity $\ovR$ as a function of $L_z$ at the Solar azimuth ($\phib - \phisun = 30^\circ$). The velocity is sampled from particles within a narrow slice ($|\Delta\varphi| < 0.5^\circ$) centred on the Sun. The general relationship between $\ovR$ and $L_z$ can be understood from the orientation (aligned or anti-aligned with the bar) and the rotating direction (prograde or retrograde with respect to the bar) of the closed parent orbits: e.g. orbits outside the OLR are aligned with the bar and are retrograding so at the Solar azimuth the closed orbit points inwards (i.e. $v_R < 0$). Within the epicycle approximation, the mean radial velocity at Solar azimuth is, by differentiating and averaging \eqref{eq:nonres_sol_const_R} with respect to time,
\begin{equation}
  \ovR = \left[ \frac{2 \Omega \Phim}{R} + \left(\Omega - \Omegap\right) \frac{\partial \Phim}{\partial R} \right]_{\Rg} \frac{m \sin m \left(\phisun - \phib \right)}{\kappa^2 - m^2 \left(\Omega - \Omegap\right)^2},
  \label{eq:mean_vR}
\end{equation}
where $\Rg = \Lz / \vc$, $\Omega = \vc / \Rg$, and $\kappa = \sqrt{2}\Omega$. The above equation delivers the black line in \Figref{fig:LzRvR_a10_wb40_s40}, which qualitatively explains the numerical results. The positive peak just behind the OLR line was the original interpretation for the Hyades stream by \cite{kalnajs1991pattern} and the Hercules stream by \cite{dehnen2000effect}. The small positive peak at CR is due to the resonantly trapped orbits reaching the Solar azimuth as they rotate around the Lagrange point ${\rm L}_{4,5}$. \cite{perez2017revisiting} attributed the Hercules stream to this peak, and supported the idea of a slow/long bar. The dotted blue curve in \Figref{fig:LzRvR_a10_wb40_s40} shows the result after imposing the \textit{Gaia} selection function, which deviates from the non-biased result mostly at small $\Lz$.

\begin{figure}
  \begin{center}
    \includegraphics[width=8.3cm]{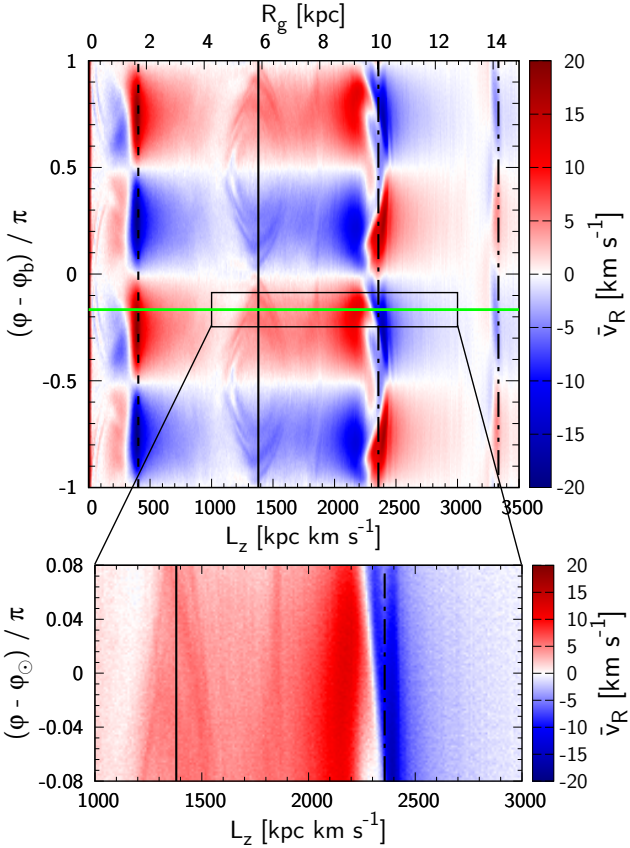}
    \vspace{-2mm}
    \caption{Azimuthal dependance of $\ovR$ vs. $L_z$. Black lines show the $\Lz$ of ILR (dotted), CR (solid), OLR (dot-dashed), and O11R (dot-dot-dashed ) at $\JR=0$, respectively. Green line indicates the azimuth of the Sun and the rectangle window indicates the range of \textit{Gaia} DR2. Bottom panel magnifies the \textit{Gaia} region and applies the spatial selection bias (eq.~\ref{eq:selection_function}).}
    \label{fig:const_bar_LzphivR}
  \end{center}
\end{figure}

One of the central benefits from \textit{Gaia} is that we can now observe stars over a wide range in Galactic azimuth, and this dependence was quantified by \cite{friske2019more} and is also shown in \Figref{fig:GaiaDR2} (b). Analogously, we show in \Figref{fig:const_bar_LzphivR} the $\varphi$-$\Lz$ dependence of the mean radial velocity $\ovR$. The vertical black lines show, from left to right, the ILR (dotted), CR (solid), OLR (dot-dashed) and 1:1 resonance (dashed) at $\JR=0$. The findings agree with previous studies: the sign of $\vR$ flips when we pass through the bar's major/minor axes, and through ILR/OLR. We also see a weak eye-like shape of orbits trapped in the CR rotating around the Lagrange points at $\varphi - \phibar = \pm \pi/2$. The bottom panel zooms into the \textit{Gaia} DR2 area (marked by a black box) and applies the spatial selection to compare with \Figref{fig:GaiaDR2} (b). We see a pair of positive and negative stripes at the OLR and a broad stripe at the CR that narrows towards the bar major axis. This somewhat resembles the \textit{Gaia} data but is far from qualitatively matching the full pattern.

\begin{figure}
  \begin{center}
    \includegraphics[width=8.5cm]{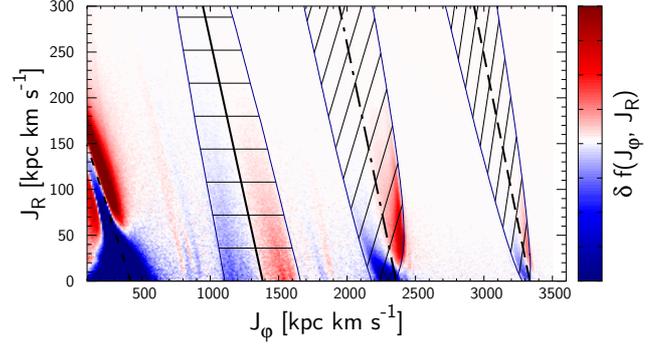}
    \vspace{-8mm}
    \caption{Change in action distribution $\delta f(\Jphi,\JR)$ from the initial unperturbed state. Densities change when trapped orbits librate between regions of different initial phase-space density around the resonance lines (thick black lines) along contours of constant $\Jf$ (thin black lines). Purple curves mark the maximum libration range.}
    \label{fig:J_df}
  \end{center}
\end{figure}

\Figref{fig:J_df} shows the change in action distribution after the bar has fully developed. Orbits near the resonances become trapped and librate back and forth across the resonance along constant $\Jf$ (thin black rungs), but not further than the purple boundary (eq.~\ref{eq:DJmax}). At the CR, $\JR$ is conserved, so the exponential disc profile in $\Jphi$ implies a mild redistribution from small to large $\Jphi$. The other resonances redistribute towards larger $\JR$, with extreme effect at the ILR, where the resonance line almost coincides with constant $\Jf$.

\begin{figure}
  \begin{center}
    \includegraphics[width=8.5cm]{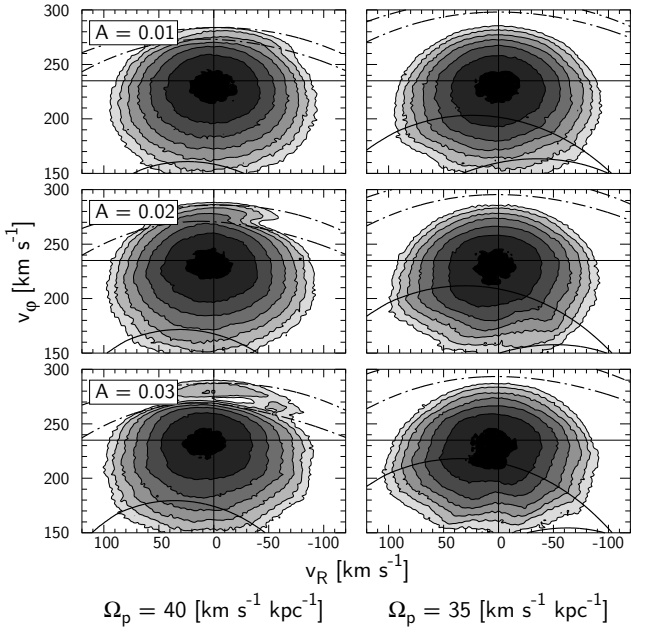}
    \vspace{-7mm}
    \caption{Velocity distribution $f(\vR,\vphi)$ in the Solar neighborhood (distance from Sun $s < 0.3 \kpc$) perturbed by a steadily rotating bar. The solid and dashed curves mark the separatrices ($\Ep = \Epsep$) of the CR and the OLR. The contours are logarithmic with a 0.44 dex spacing.}
    \label{fig:const_bar_vRvphi}
  \end{center}
\end{figure}

Finally, \Figref{fig:const_bar_vRvphi} shows the velocity distribution $f(\vR,\vphi)$ in the Solar neighborhood ($s < 0.3 \kpc$) drawn from test particle simulations with three different bar strength (row) and two different pattern speeds (columns) both identified as a slow bar. A wider range of pattern speeds, including the traditional fast bar, will be shown with their corresponding slowing bar models in the next section. The black solid and dot-dashed curves mark the separatrices enclosing the resonant regions of the CR and the OLR as done in \cite{monari2017distribution}. At $\Omegap = 40 \kmskpc$, orbits trapped by the OLR appear distinctively as an arch at $\vphi \sim 280 \kms$, whereas the velocity distribution of orbits trapped by the CR shows little contrast to the surrounding non-resonant region. For this reason, past studies disfavoured linking the Hercules stream with the CR \citep[][]{dehnen2000effect, monari2016staying, fragkoudi2019ridges}. In the next section, we will show that this problem is naturally resolved by a decelerating bar where the CR captures more stars and on a different action distribution as it form further inside and then sweep outwards.

However we mention here again that studies by \cite{perez2017revisiting,Monari2019signatures} report better success in reproducing the Hercules stream with a constantly rotating bar which suggests that higher-order modes of the bar can be important for shaping the distinct outline of the Hercules stream. We further note that the Hercules stream can also be reproduced by transient spirals alone or in combination with the bar \citep[]{Hunt2018Transient}. Since none of these effects are ignorable, it will be important in the future to combine these models and distinguish their role, once the kinematic consequence of individual perturbations are well understood.

\subsection{Slowing bar}
\label{sec:slowing_bar}

We now consider the effect of a slowing bar. Two new processes arise from resonance sweeping: dragging of resonantly trapped orbits along with the moving resonance and trapping of non-resonant orbits when the resonance crosses over their domain. At current stage, capture and loss from resonances require numerical treatment. Nevertheless the analytic approach predicts that resonance volume shrink with decreasing amplitude $A$ and increasing slow-down rate $\eta$ of the bar, so we have a naive, but firm expectation that the capturing and retention rates will also decrease with increasing $\eta/A$.

\begin{figure}
  \begin{center}
    \includegraphics[width=7.5cm]{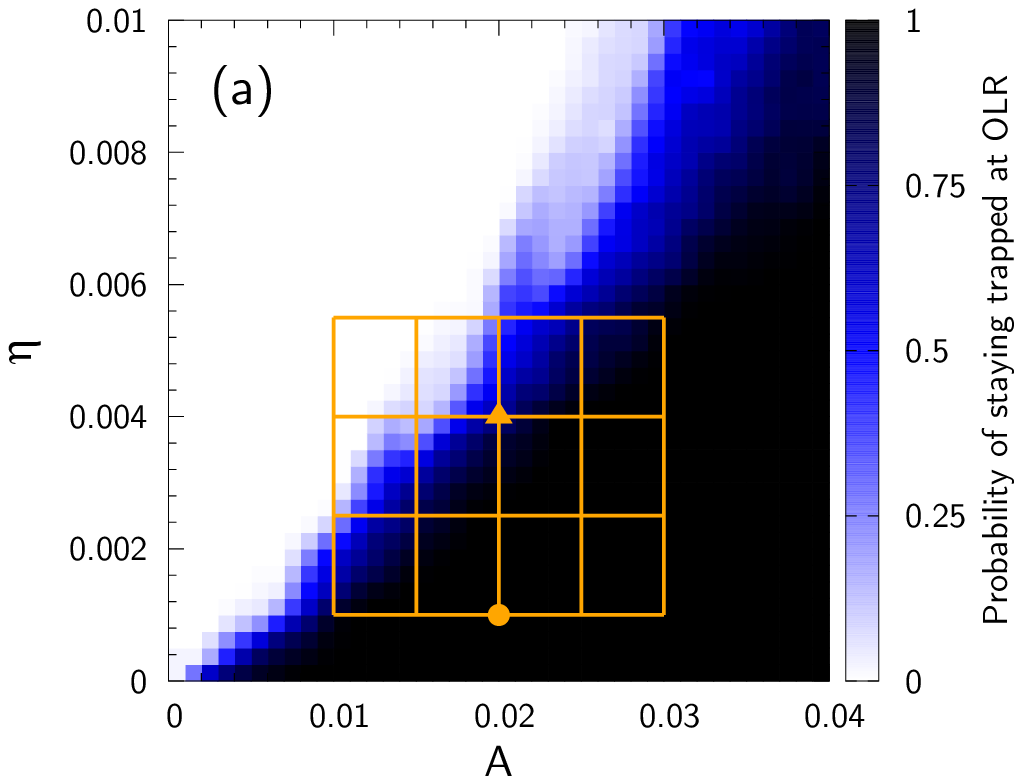}
    \includegraphics[width=7.5cm]{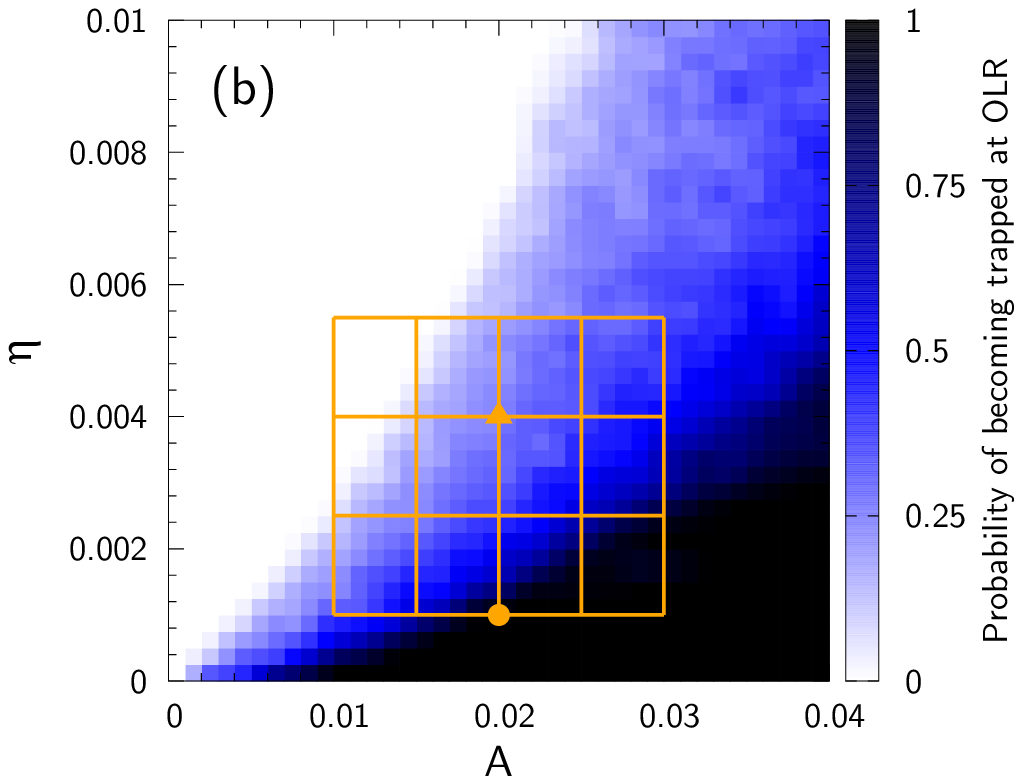}
    \vspace{-2mm}
    \caption{Top (a): Fraction of orbits initially trapped at OLR being dragged by the slowing bar as a function of bar strength $A$ and slowing rate $\eta$. The orange circle and triangle mark the parameters of our standard slowly and rapidly decelerating bar. Results using the full grid are shown in \Figref{fig:vRvphi_A_e}. Bottom (b): Fraction of orbits \textit{becoming} trapped by the OLR. The blue transition zone is broad since resonant capturing depends on the initial angles.}
    \label{fig:res_frac_ae}
  \end{center}
\end{figure}

In \Figref{fig:res_frac_ae}~(a) we show the probability of being successfully dragged by the moving OLR as a function of bar strength $A$ and slowing rate $\eta$. The orbits are initially trapped in the OLR at $\Omegap = 60 \kmskpc$ and the bar is subsequently slowed down to $\Omegap = 30 \kmskpc$. The plot shows the fraction of successfully dragged stars defined as those originally in the OLR that then experience a relative increase $\delta\Lz/\Lzo > 0.2$. The threshold is chosen to safely exceed the maximum libration amplitude in $\Lz$. For each parameter set $(A, \eta)$, we use 100 particles with the same initial actions $(\JR, \Jphi) = (19.4, 1524.5) \kpc\kms$ but with random angular phase. The initial actions are placed exactly on the resonance line to ensure trapping independent of angular phase. The result confirms the analytical expectation that the retention probability increases with $A$ and decreases with $\eta$. The critical boundary is fairly linear which backs our idea that $\eta/A$ satisfactorily describes the dragging efficiency. Similarly, \Figref{fig:res_frac_ae}~(b) shows the capturing rate by the moving OLR. Here we set the initial actions $(\JR, \Jphi) = (19.4, 1921.1) \kpckms$ which is outside the resonance. The blue transition region roughly matches that of \Figref{fig:res_frac_ae}~(a), but is wider, indicating a strong dependence of resonant capturing on angular phase.

Figure \ref{fig:res_frac_ae} implies three parameter regimes in which the dynamical consequence of a slowing bar differs qualitatively: In the white regime, the resonance sweeping is too fast or the bar is too weak for resonant trapping to occur. The extreme opposite is the black regime where most orbits are captured and dragged by the resonance resulting in a great migration of orbits. In the blue intermediate regime, most resonant orbits are dragged but not all non-resonant orbits are captured. This discussion neglects that orbits can also be kicked out of and into resonances by gravitational fluctuations due to satellite galaxies, transient spiral arms, and giant molecular clouds, a discussion which we defer to a later study. The CR shows similar behaviour to the OLR, but the situation is more complicated due to the series of small higher-order resonances piling up towards the CR, leading to chaotic behaviour. We therefore prefer the parameter map at OLR to qualify the slowing regime, while using the CR as a corroborating source of evidence. In the following, we discuss results from the full orange parameter grid, which covers all three regimes, first starting with an in-depth analysis of the slowly (orange circle) and the rapidly (triangle) decelerating bar.

\subsubsection{Slowly decelerating bar}
\label{sec:slowly_slowing_bar}

\begin{figure*}
  \begin{center}
    \includegraphics[width=15.5cm]{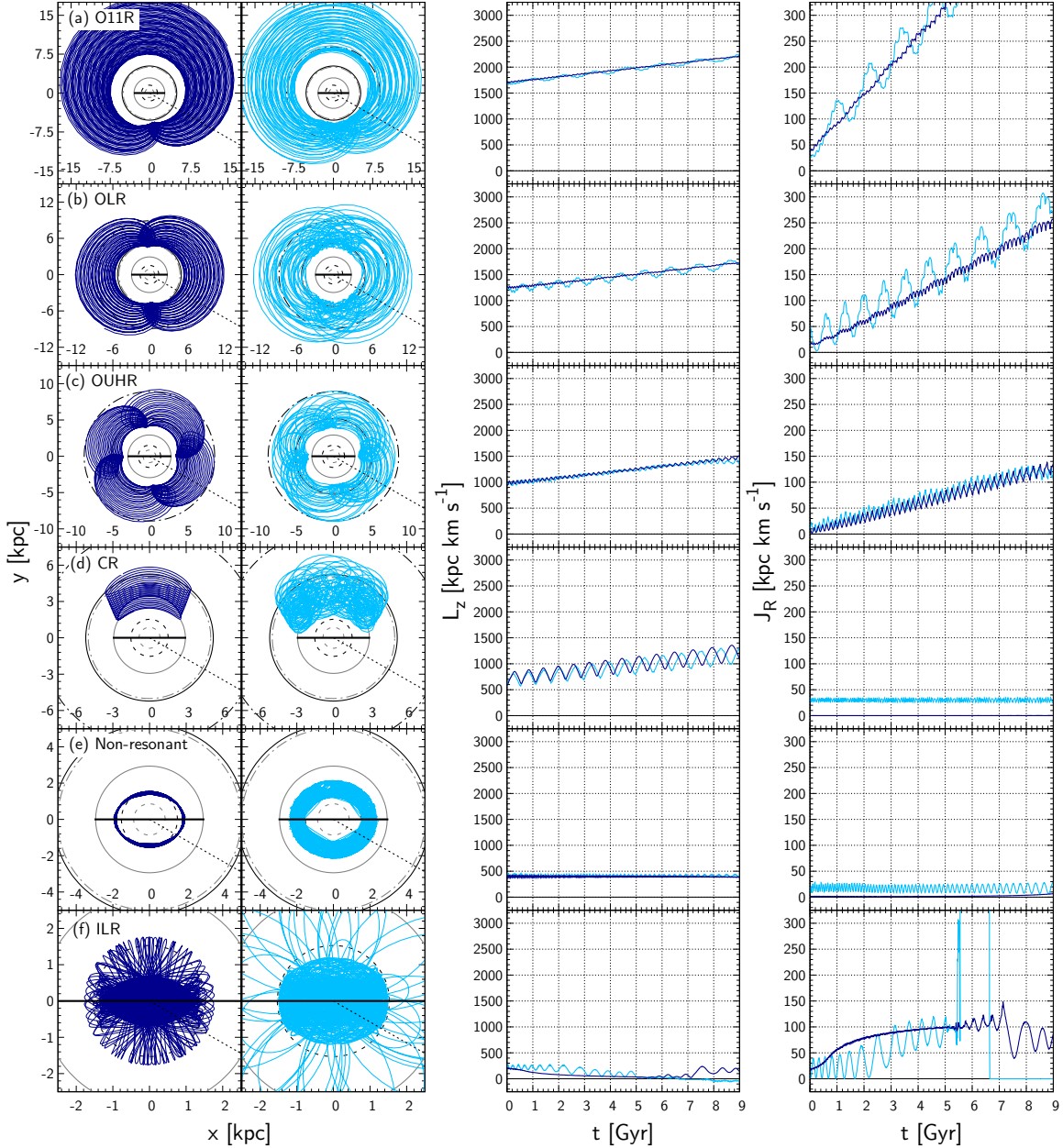}
    \vspace{-4mm}
    \caption{Orbits swept by the resonance of a slowing bar decreasing its pattern speed from $\Omegap = 80$ to 45 $\kmskpc$ in 9 $\Gyr$ ($\eta = 0.001$). Each row displays the trajectory of a single star gradually changing its orbit. Dark blue orbits (left column) are initially closed while light blue orbits (2nd column) are initially non-closed. The grey and black circles are the initial and final radii of ILR (dashed), CR (solid), and OLR (dot-dashed). Drift in angular momentum (3rd column) is only seen for resonantly trapped orbits. The radial action (right column) increases when dragged at all resonances except at the CR.}
    \label{fig:slow_orbit_Lz_JR}
  \end{center}
\end{figure*}

\Figref{fig:slow_orbit_Lz_JR} analyses typical orbits in a slowly decelerating bar ($\eta = 0.001$, $A = 0.02$, and thus $\eta/A = 0.05$), where $\Omegap$ decreases from 80 to 45 $\kmskpc$ in 9 $\Gyr$ (with a transition period of $t_2 - t_1 = 1 \Gyr$). As in \Figref{fig:orbitRF_a10_wb40}, we show more eccentric orbits (light blue, second column) and their closed parents (dark blue, first column). Grey and black circles indicate the initial and final resonance radii for ILR (dotted), CR (solid), and OLR (dot-dashed). The orbits have initial guiding radii of (a) 7.1, (b) 5.2, (c) 4.0, (d) 3.0, (e) 1.7, and (f) 1.0 kpc. The other columns provide the evolution of the actions.

The rows (a)-(d) show orbits trapped and dragged outwards by O11R, OLR, OUHR, and CR. In the rotating frame of the bar, frame deceleration causes an Euler force $\dot{\bm \Omega}_\mathrm{p}\times{\bm R}$ responsible for the slight anti-clockwise turn of the orbits' configuration. The orbit in row (e) remains non-resonant and thus roughly maintains its orbital configuration. The orbit (f) is dragged by the ILR until it turns chaotic. The qualitative behaviour of the actions agrees with the secular perturbation theory (Section \ref{sec:resonant_dragging}): at the outer resonances (a)-(c), both $L_z$ and $\JR$ continuously increase, whereas at the CR (d), only $L_z$ enhances while $\JR$ is kept fixed; actions of non-resonant orbits (e) are unchanged; at the ILR (f), $L_z$ declines while $\JR$ rises. 

\begin{figure*}
  \begin{center}
  \includegraphics[width=17.5cm]{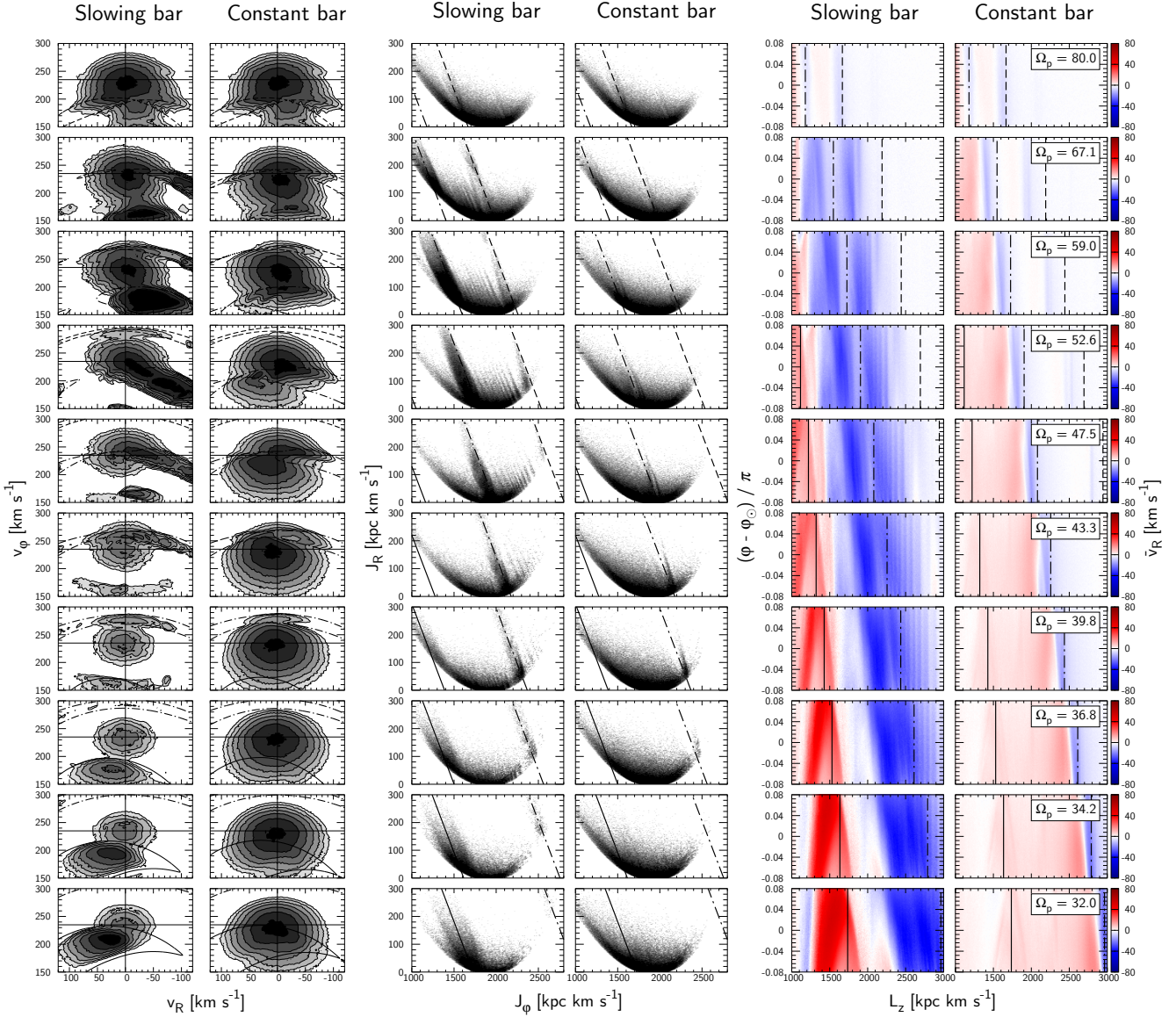}
    \vspace{-2mm}
    \caption{Simulated phase space when the bar decelerates slowly from $\Omegap = 80$ to 45 $\kmskpc$ in 9 $\Gyr$. Time-interval between the rows is $2 \Gyr$ (last three snapshots extends beyond the age of the Galaxy). Each pair of columns compare the decelerating bar vs. a constantly rotating bar at the same $\Omegap$ indicated by white panels in the right-hand column. Black solid, dot-dashed, and dashed lines represent CR, OLR, and O11R respectively. In the velocity plane we mark the separatrix of the resonances, while in the action plane we draw the exact resonance line, and in the $(L_z, \varphi)$ plane we indicate the loci of resonances at $\JR = 0$. The OLR and CR capture and retain the majority of orbits along their way resulting in a unrealistic intense stellar stream.}
    \label{fig:all_A20_e10}
  \end{center}
\end{figure*}

\Figref{fig:all_A20_e10} follows the evolution of the phase-space distribution perturbed by a slowly decelerating bar. The rows from top to bottom show the distribution every $2 \Gyr$, denoting the pattern speed $\Omegap$ on the right in $\kmskpc$. We provide the velocity distribution and action distribution near the Sun ($s < 0.3 \kpc$), as well as the mean radial velocity $\ovR$ in the $\Lz$-$\varphi$ plane in a narrow slice around the Solar azimuth applying our approximated \textit{Gaia} selection function. For each panel we provide to its right the comparison case of a constantly rotating bar with identical amplitude and current $\Omegap$.

The local velocity plane (left-hand columns) is dominated by the main resonances; first, the OLR captures and carries away the majority of non-resonant orbits leaving behind a significantly depleted phase space, until the CR brings along the next swath of stars. At around $\Omegap = 43.3 \kmskpc$, an arch opened towards high $\vphi$ develops below the circular velocity. Orbits below this arc typically have sufficient kinetic energy to cross over the crest of the effective potential $\Phi - \frac{1}{2} \Omegap^2 R^2$ in the rotating frame and thereby wander in and out the bar regime. \cite{fux2001order} proposed that the Hercules stream may be associated with these orbits. However, this slowly decelerating bar here will always have radically too strong resonance occupation to match the \textit{Gaia} data.

The action plane (middle columns) shows similarly the capture and drag by the main resonances. In between the OLR and the O11R, we see multiple narrow lines. We confirm that they are due to orbits trapped and dragged by minor resonances (e.g. 2:3 resonance). The occupation on minor resonances depends on their stability under the deceleration but also on depletion by anteceding resonances; minor resonances behind the OLR are less prominent since the OLR sweeps away most of the non-resonant orbits in advance.

In the right columns of \Figref{fig:all_A20_e10}, the amplitude of $\ovR(\Lz,\varphi)$ near resonances continues to increase as they keep collecting stars. The red spear-like stripe at the CR increases in strength at the edge indicating accumulation of orbits near the separatrix. The blue stripe associated with the OLR widens to the left over time since trapped orbits increase $\JR$ while dragged and thus satisfy the resonance condition at a relatively lower $\Lz$ compared to those with small $\JR$.

\subsubsection{Rapidly decelerating bar}
\label{sec:rapidly_slowing_bar}

\begin{figure*}
  \begin{center}
    \includegraphics[width=17.5cm]{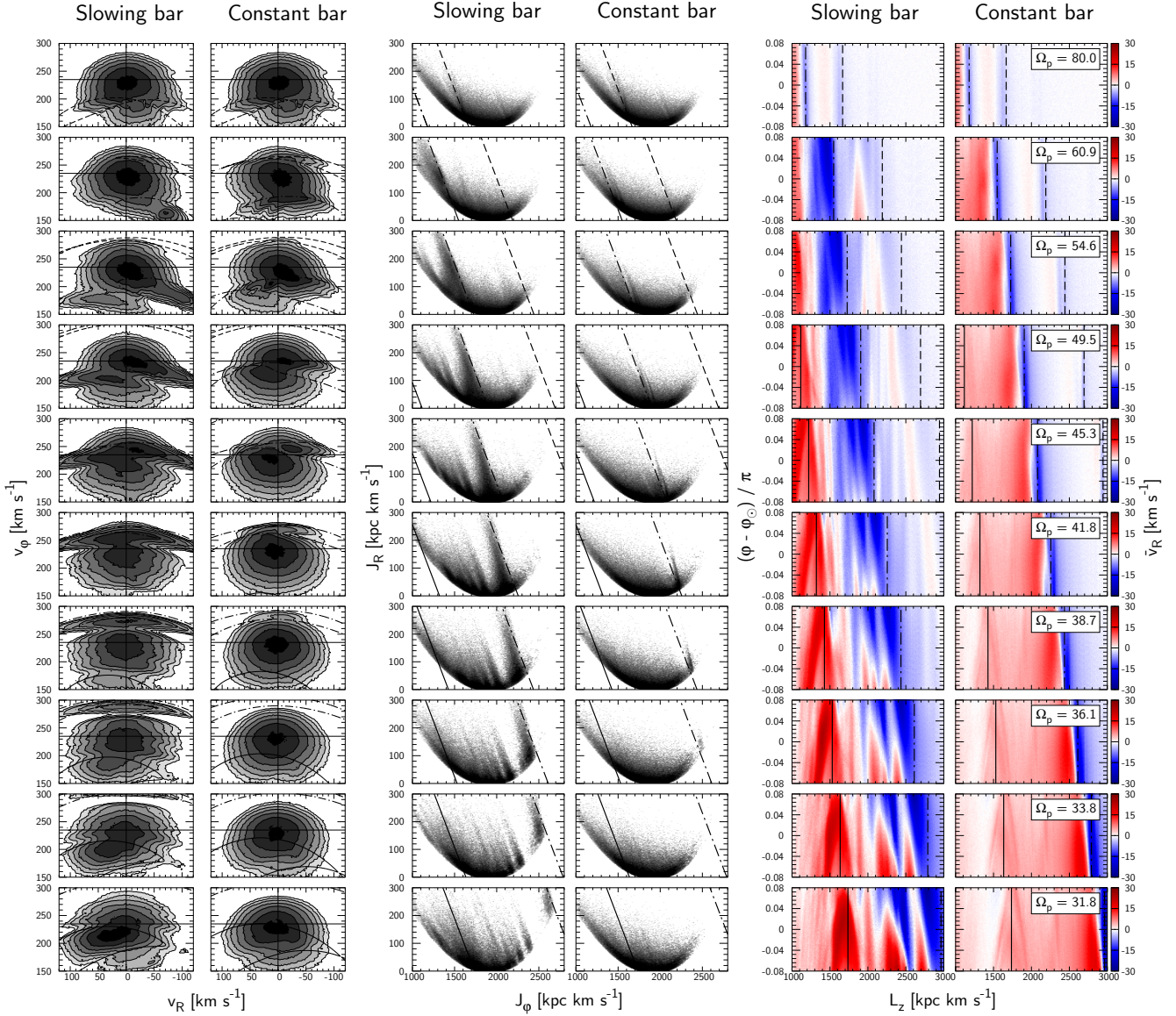}
    \vspace{-2mm}
    \caption{Simulated phase space as in \Figref{fig:all_A20_e10} plotted every $0.56 \Gyr$ when the bar decelerates rapidly from $\Omegap = 80$ to $30 \kmskpc$ in $5.6 \Gyr$. Unlike the case of a slowly slowing bar, the effect of resonance sweeping is moderate, yet enhances the resonance features compared to a constantly rotating bar (right of each column). The CR reproduce well the Hercules stream in the velocity plane and the spear-like double peak in $\ovR(\Lz,\varphi)$ seen in the \textit{Gaia} data (\Figref{fig:GaiaDR2}).}
    \label{fig:all_A20_e40}
  \end{center}
\end{figure*}

N-body studies indicate that bars slow down more rapidly than we have assumed in Section \ref{sec:slowly_slowing_bar}. In accordance with \cite{aumer2015origin} we choose the slowing rate $\eta = 0.004$ as indicated by the orange triangle in Fig.\ref{fig:res_frac_ae}. The pattern speed decreases from $\Omegap = 80$ to $30 \kmskpc$ in $5.6 \Gyr$. The strength of the bar is unchanged ($A = 0.02$), so $\eta/A = 0.2$. \Figref{fig:all_A20_e40} shows from left to right the velocity distribution, the action distribution, and the mean $\vR$ in the $\Lz$-$\varphi$ plane, along with the results of a constantly rotating bar on the right of each column.

In the velocity plane, the resonances are much smaller in volume than those of the constantly rotating bar, but appear more distinctively. At around $\Omegap = 36.1 \kmskpc$, the orbits trapped at the CR form a peak that resembles the observed Hercules stream much better both in strength (by sweeping up more stars) and in location (due to the shrinkage of the resonance region towards high $\vR$): The Hercules stream modeled by a constantly rotating bar \citep[see also][]{perez2017revisiting} is far too symmetric in $\vR$, while the decelerating bar provides the strong asymmetry which previously could only be achieved by the OLR \citep[e.g.][]{dehnen2000effect}.

In the action plane, multiple inclined ridges appear between the CR and the OLR; the small capturing rate at the OLR leaves opportunity for orbits to be captured into the minor resonances passing later. Obviously the signatures of minor resonances will be enhanced if we add higher order modes of the bar. These inclined ridges are also seen in the \textit{Gaia} data (\Figref{fig:GaiaDR2} (a)). However the ILR of spiral arms also lies in the vicinity \citep{Sellwood2019Discriminating} making the exact identification of individual ridges difficult.

The right columns in \Figref{fig:all_A20_e40} show that the amplitude in the $\ovR$ increases much less than for the slowly decelerating bar due to the smaller capturing rates (note the different colour scales). Again at the CR, the newly captured orbits accumulate near the resonance boundary and form a spear-like (an eye-like, if we saw the full $\varphi$ range; \Figref{fig:LzphivR_a10_e40}) structure which closely resembles the double positive peak in the \textit{Gaia} data. In between the CR and the OLR line, we observe roughly two pairs of positive and negative stripes.

\begin{figure}
  \begin{center}
    \includegraphics[width=8cm]{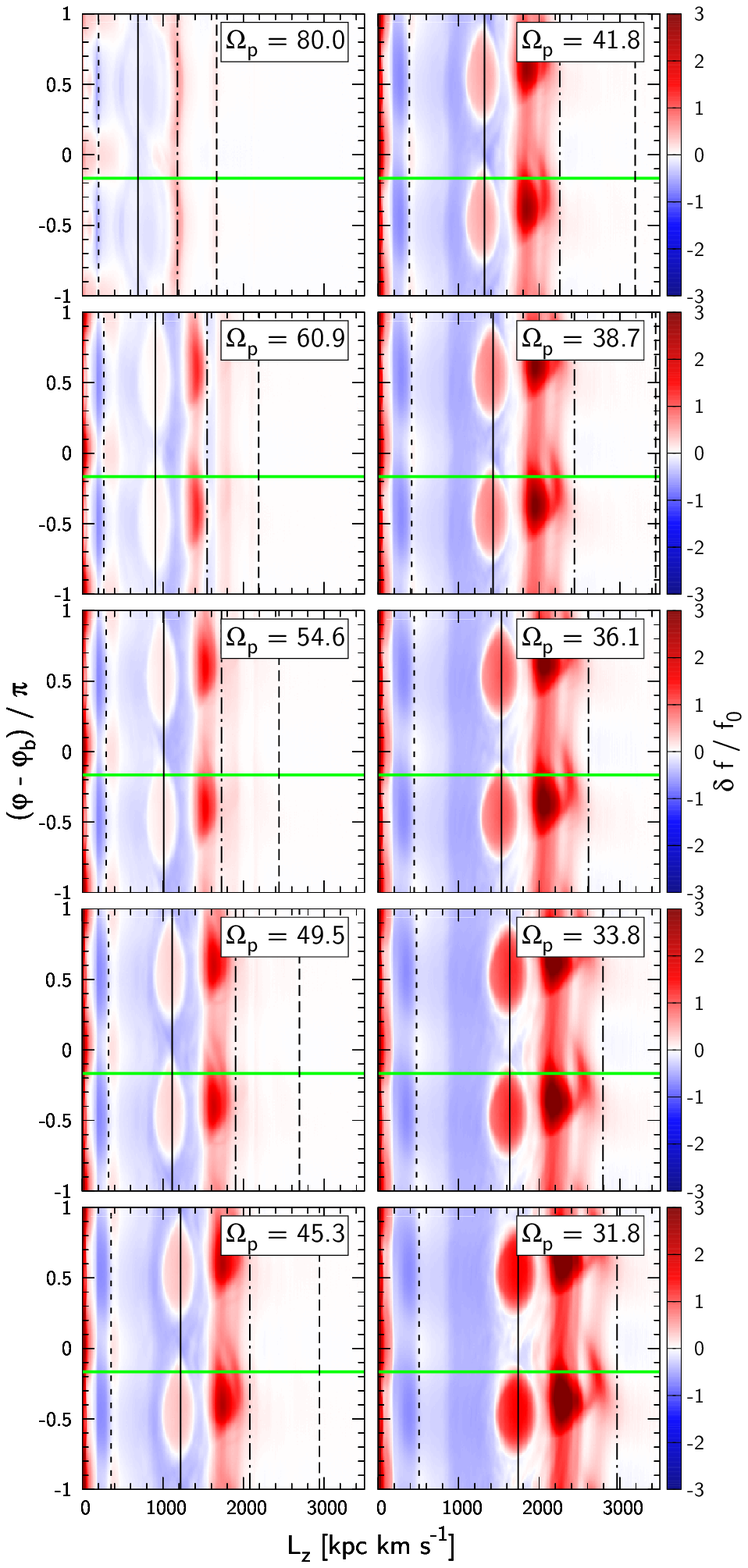}
    \vspace{-4mm}
    \caption{Evolution of the distribution $\delta f(\Lz, \varphi)$ from the unperturbed state $f_0$ when the bar slows rapidly from $\Omegap = 80$ to $30 \kmskpc$ in $5.6 \Gyr$. By definition, $\delta f / f_0 > -1$. The black lines represents, from left to right, ILR, CR, OLR, and O11R at $\JR=0$. The green line marks the Solar azimuth. Stars captured at the CR co-move with the resonant line while stars resonating at the OLR fall behind in $\Lz$ due to their increase in $\JR$.}
    \label{fig:Lzphidf_a10_e40}
  \end{center}
\end{figure}

\begin{figure}
  \begin{center}
    \includegraphics[width=8cm]{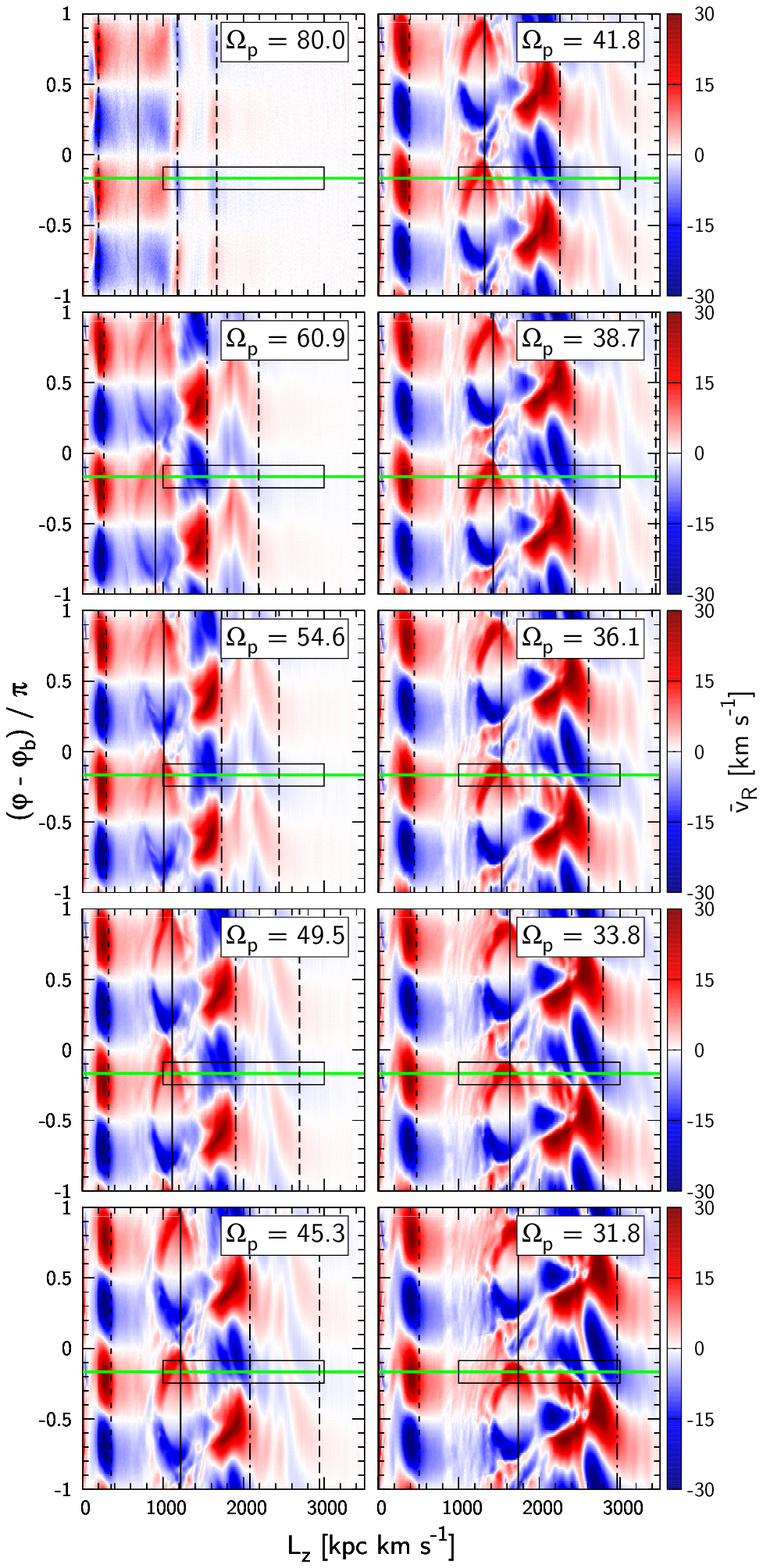}
    \vspace{-4mm}
    \caption{The mean radial velocity distribution $\ovR(\Lz, \varphi)$ perturbed by a rapidly slowing bar. The black narrow rectangles represent the range of \textit{Gaia} data which we used to show results in \Figref{fig:all_A20_e40}. Behind the OLR line (dot-dashed), two negative (blue) peaks are formed near Solar azimuth indicated by the green horizontal line. Comparison with \Figref{fig:Lzphidf_a10_e40} implies that the inner peaks are formed by the resonant orbits dragged and heated by the OLR.}
    \label{fig:LzphivR_a10_e40}
  \end{center}
\end{figure}

To better understand the origin of the multiple stripes in $\ovR(\Lz,\varphi)$, we show in \Figref{fig:Lzphidf_a10_e40} the relative change of number density with respect to the initial distribution $(\delta f/f_0 = f/f_0 - 1)$, and in \Figref{fig:LzphivR_a10_e40}, the $\ovR$ over the full $\varphi$ range without imposing the selection bias (indicating the Solar azimuth $\phib - \phisun = 30^\circ$ with a green line). \Figref{fig:Lzphidf_a10_e40} shows that, at the CR, trapped stars follow the resonance line plotted for $\JR = 0$, whereas at the OLR, orbits lag increasingly behind the $\JR = 0$ resonant line since their $\JR$ increases (resonance line is negatively inclined in $\JR$ vs. $\Lz$ as in \Figref{fig:J_HJ}). Blue/Underpopulated areas are left behind the travelling OLR and CR, as these resonances sweep part of stars along their path. An even more intense depopulation is caused by the ILR where trapped stars drift towards lower $\Lz$ and larger $\JR$. Note that the initial distribution in $\Lz$ declines exponentially so a group of orbits moving toward high $\Lz$ with constant number density show apparent enhancement in $\delta f$. The mean radial velocity shown in \Figref{fig:LzphivR_a10_e40} is strongly distorted and much more complex than the constantly rotating bar case shown in \Figref{fig:const_bar_LzphivR}, particularly between CR and OLR. Nevertheless we can identify two pairs of blue/red peaks behind the OLR near the Solar azimuth ($\Lz \sim$ 2400 and 2800 $\kpckms$ in the last frame). These structures appear as multiple stripes when seen in the \textit{Gaia} range indicated by the narrow rectangle. By comparing \Figref{fig:Lzphidf_a10_e40} (the location of the orbits dragged by the OLR) and \Figref{fig:LzphivR_a10_e40} (the location of the stripes) we conclude that the pair of $\ovR$ stripes just inside the OLR are due to the orbits freshly trapped by the OLR while the stripes that appear further inside the OLR line are due to the superposition of orbits dragged/heated by the OLR and orbits trapped in minor resonances. The weakly negative $\vR$ outside the OLR are associated with non-resonant $x_1$ orbits swept but not captured by the O11R.

The location and inclination of these multiple stripes at large $\Lz$ do not match perfectly with the \textit{Gaia} data (\Figref{fig:GaiaDR2} (a)). The neglected transient spiral arms can play at least two important roles here: i) Spiral patterns can independently form these stripes by leaving scars in the action distribution near resonances and also in the angular distribution, due to their non-adiabatic emergence, which develops into multiple fine stripes in $\ovR$ as they phase mix over time \citep{Hunt2019signature}. ii) Scattering by transient spiral arms would reduce the occupation particularly of the bar OLR, since it trades stars with the much less densely occupied surrounding phase space at high $\JR$. This would significantly weaken the contribution of the OLR to structures at large $\Lz$. The relative position of the resonances will also vary with the inclination of the circular speed curve: negative/positive inclination leads to smaller/larger separation of the resonances.

\subsubsection{Dependence on initial pattern speed}
\label{sec:initial_pattern_speed}

\begin{figure}
  \begin{center}
    \includegraphics[width=8.5cm]{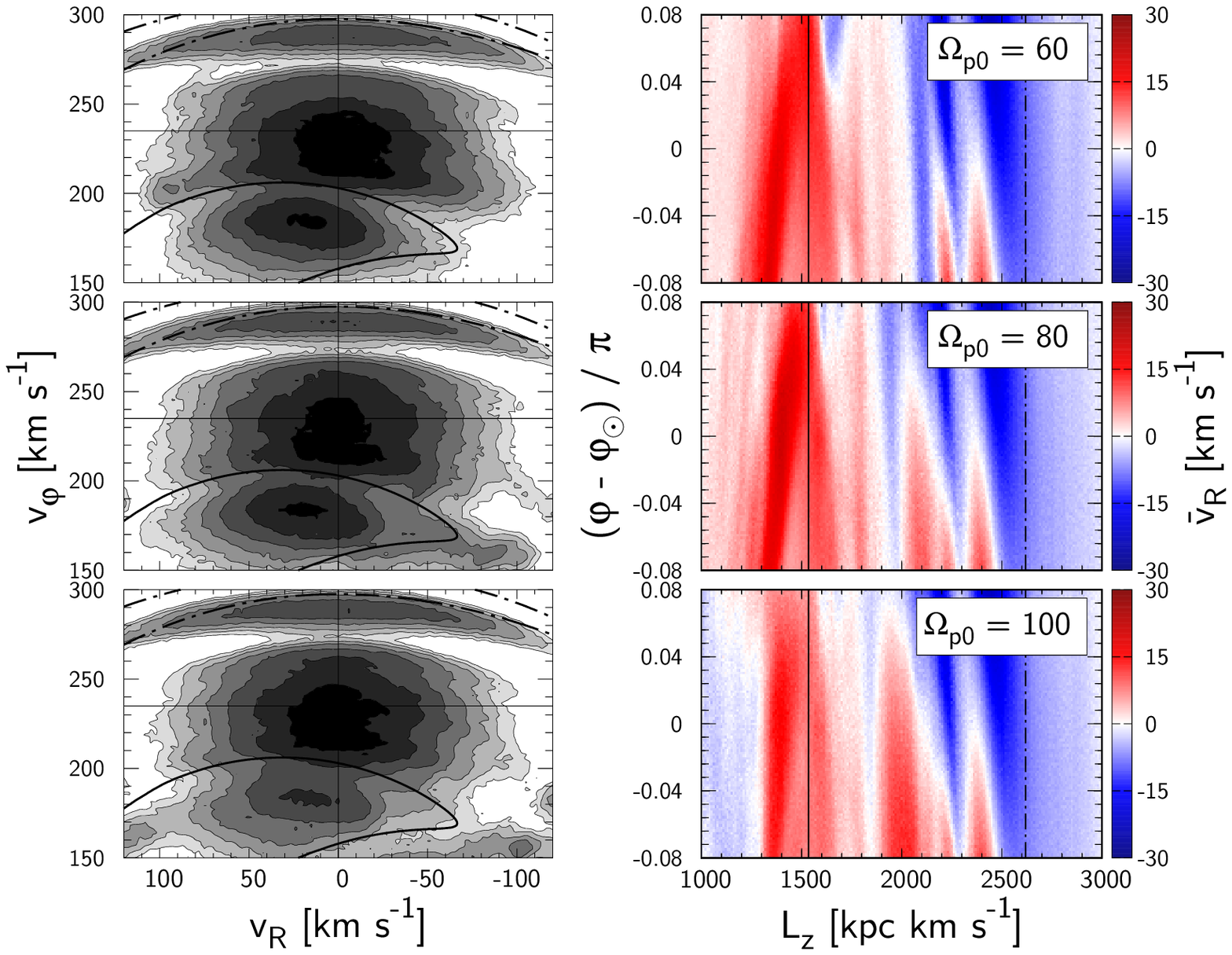}
    \vspace{-6mm}
    \caption{Dependence on initial pattern speed: from top row, $\Omegapo$ = 60, 80, and 100 $\kmskpc$. The bar is rapidly decelerating ($A = 0.02, \eta = 0.004$) and the present pattern speed is $\Omegap$ = 36 $\kmskpc$.}
    \label{fig:all_A20_e40_wb0}
  \end{center}
\end{figure}

We have so far discussed the impact of the slowing rate while keeping the initial pattern speed fixed at $\Omegapo = 80 \kmskpc$. We now take a look at the impact of $\Omegapo$. \Figref{fig:all_A20_e40_wb0} plots $\ovR(\Lz,\varphi)$ and $f(\vR,\vphi)$ for three different choices of $\Omegapo$ increasing from top to bottom. The faster the bar is originally, the further inside the disc the original locations of the resonances will be, and thus the larger the volume of phase space swept by the resonance. However, the variation of the initial pattern speed do not result in a big difference here because the capturing rate is relatively low for a rapidly decelerating bar.

\subsubsection{Determining the slowing rate of the bar}
\label{sec:determine_slowing_rate}

\begin{figure*}
  \begin{center}
    \includegraphics[width=17.5cm]{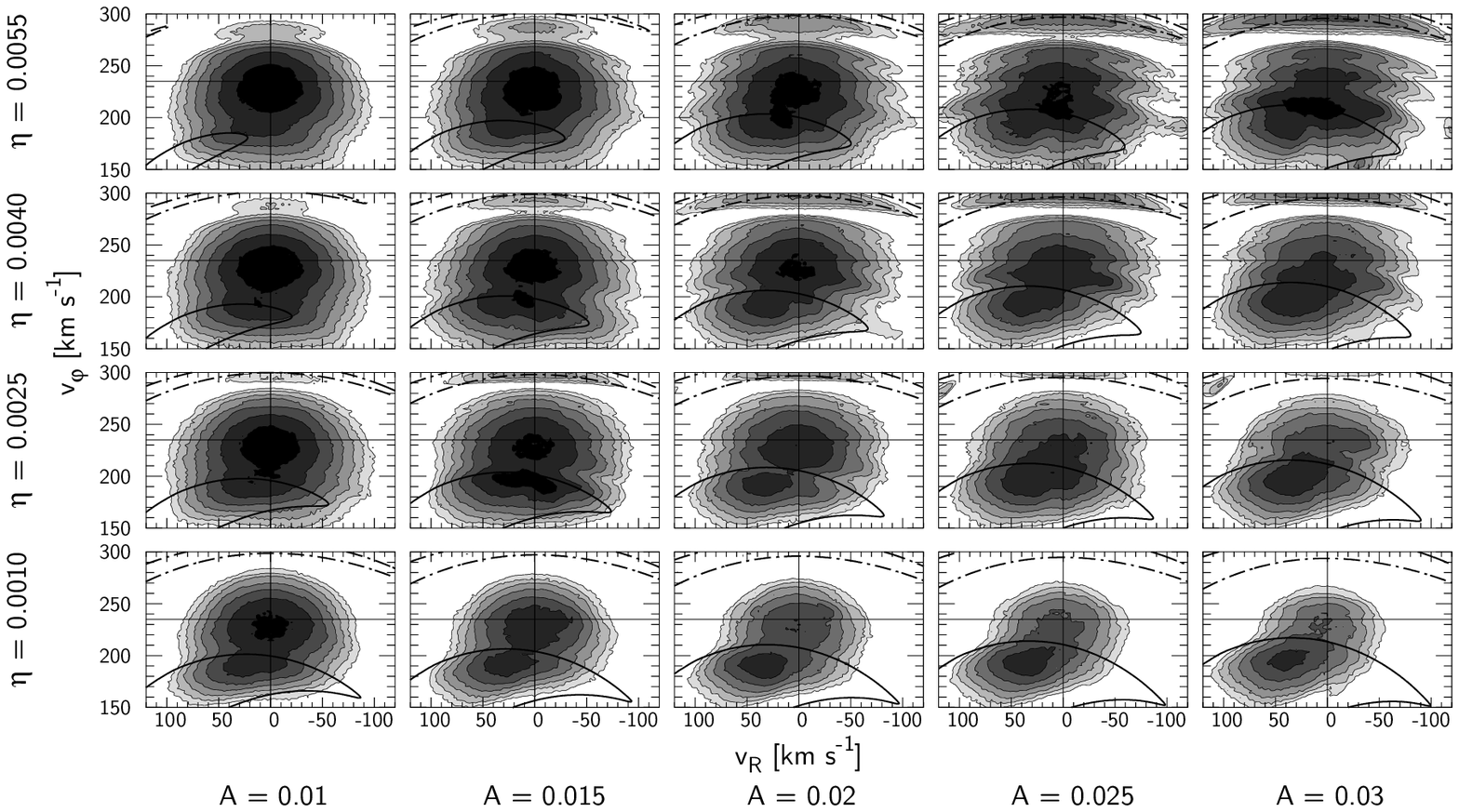}
    \vspace{-4mm}
    \caption{Simulated distributions in local velocity space for the grid of bar strength $A$ and slowing rate $\eta$ shown in \Figref{fig:res_frac_ae}.}
    \label{fig:vRvphi_A_e}
  \end{center}
\end{figure*}

\begin{figure}
  \begin{center}
    \includegraphics[width=8.5cm]{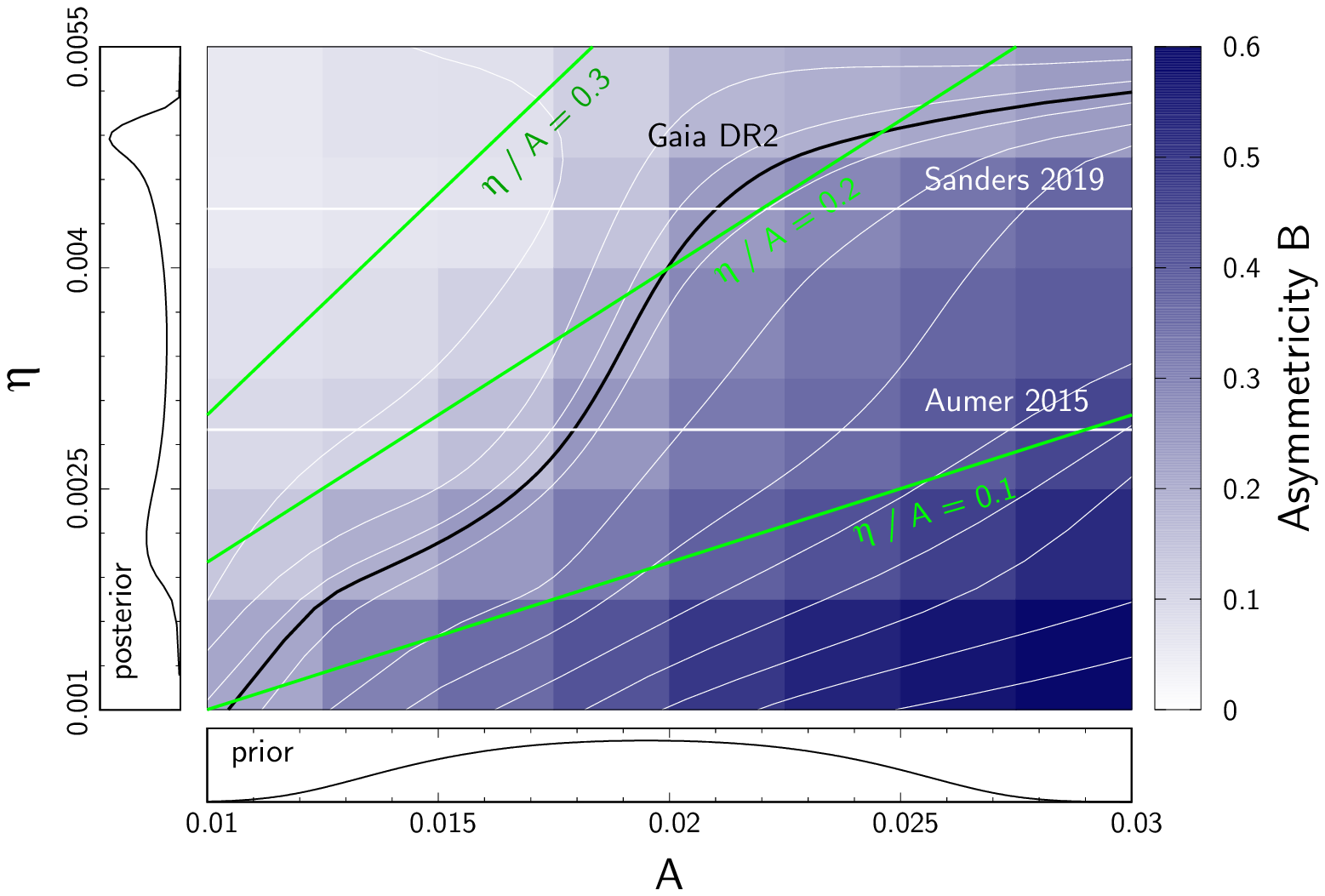}
    \vspace{-6mm}
    \caption{Asymmetricity $B$ in $\vR$ in the Hercules region (see equation \ref{eq:asymmetricity}), plotted over bar strength $A$ and slowing rate $\eta$. The colours map the simulated values, with the white contours gained from spline interpolation. The black contour marks the value obtained from \textit{Gaia} DR2. We show the resulting posterior probability distributions for $\eta$ on the left panel when using a prior distribution on $A$ (bottom panel) inferred from SBM15.}
    \label{fig:vRasym_A_eta_v210}
  \end{center}
\end{figure}

Beyond constraining the current pattern speed, modeling local kinematics perturbed by a decelerating bar yields measurement of the bar's slowing rate $\dOmegap$. \Figref{fig:vRvphi_A_e} shows the local velocity distribution of stars $f(\vvel)$ for various values of bar amplitude $A$ (columns) and slowing parameter $\eta$ (rows) corresponding to the orange grid nodes in \Figref{fig:res_frac_ae}. The present pattern speed $\Omegap = 35 \kmskpc$ is determined by matching the location of the CR with the Hercules stream. The fraction of orbits trapped in the CR increases with increasing $A$ and decreasing $\eta$ in concordance with changes in retention/capture rates (see \Figref{fig:res_frac_ae}). We now need quantifiable statistics linked to this resonance occupation: the most obvious target is the {\it asymmetricity} of Hercules in the radial velocity $\vR$. We take the difference of fraction of stars with positive vs. negative $\vR$:
\begin{equation}
   B \equiv \int^{\vphimax}_{\vphimin}d\vphi \left[ \frac{\int^{\infty}_0 d\vR f(\vvel) - \int^0_{-\infty} d\vR f(\vvel)}{\int^{\infty}_{-\infty} d\vR f(\vvel)} \right].
  \label{eq:asymmetricity}
\end{equation}
We choose $\vphimin = 100 \kms$ and $\vphimax = 210 \kms$ to approximately cover the CR. \Figref{fig:vRasym_A_eta_v210} shows the asymmetricity $B$ evaluated on the parameter grid. The white contours of $B$ are gained from spline interpolation and the black contour marks the observed value ($B=0.225$) from \textit{Gaia} DR2 (\Figref{fig:GaiaDR2} (a)). $B$ increases towards large $A$ and small $\eta$, and closely follows $\eta/A$ marked by the green lines.

To estimate the posterior distribution of $\eta$ and $\eta/A$ by comparing $B$ between simulations and observations, we need an appropriate prior on $A$. SBM15 compared their gas dynamic models to the inner Milky Way photometry and constrained the possible range of their bar amplitude parameter $A_{\rm s} \in [0.4, 0.8]$. To translate their $A_{\rm s}$ to our $A$, we must correct for the disparity between the models at large radius evident in Fig.~\ref{fig:Phi2}. Since we need consistent perturbation strength near the CR, and not in the the inner bar region, we fit our model at radii beyond half of $\RCR$ which yields $A \in [0.013,0.026]$. To ensure a smooth prior obeying Cromwell’s rule, we prescribe a normal distribution $P(A)$ with mean $\mu_A = 0.0195$ and standard deviation $\sigma_A = 0.0065$ within the allowed region, and a smooth cutoff at $A = \mu_A \pm \sigma_A$:
\begin{align}
  &P(A) \propto \frac{\exp{\left(- x^2 / 2 \right)}}{\frac{1}{2}\left(\exp{|x^k|} + 1\right)}, ~~~ x \equiv \frac{A-\mu_A}{\sigma_A}
  \label{eq:prior_A}
\end{align}
where $k$ controls the steepness of the cutoff. We choose $k = 4$. The prior $P(A)$ is shown at the bottom panel of Fig.~\ref{fig:vRasym_A_eta_v210}. The posterior $P(\eta)$ is the product of the prior $P(A)$ and the slope of the observational curve $A(\eta)$ (the black curve in Fig.~\ref{fig:vRasym_A_eta_v210}):
\begin{align}
  P(\eta) = P(A(\eta)) ~ \frac{dA}{d\eta}.
  \label{eq:posterior_eta}
\end{align}
The posterior $P(\eta)$, shown in the left panel of Fig.~\ref{fig:vRasym_A_eta_v210}, has expectation value $\langle\eta\rangle = 0.0036$, median $\tilde{\eta} = 0.0039$, and standard deviation $\sigma_\eta = 0.0011$. The estimators are robust, i.e. vary by less than $10\%$ when extending the upper limits of the $\vphi$ sampling range from $\vphimax = 170 \kms$ to $220 \kms$. Decrease in the steepness of the prior from $k=4$ to $k=2$ increases the standard deviation by $2.3\%$ although merely changes the mean value of $\eta$ (variation less than $0.2\%$). The estimated bar slowing rate $\eta$ roughly agrees with values encountered in N-body simulations (horizontal white lines in Fig.~\ref{fig:vRasym_A_eta_v210}): $\eta = 0.0029$ from Fig.~2 of \cite{aumer2015origin}, and $\eta = 0.0044$ from Fig.~7 of \cite{Sanders2019pattern}. Using a current pattern speed $\Omegap = 35 \kmskpc$, our $\eta$ estimate translates to: $\dOmegap = - \eta \Omegap^2 = (- 4.5 \pm 1.4) \kms\kpc^{-1}\Gyr^{-1}$. The posterior expectation value for the ratio $\eta/A = 0.18 \pm 0.03$ is more tightly constrained as expected. These estimates agree with the visual inspection/comparison of the velocity plane.

We caution however that this analysis is based on a pure $m=2$ slowing bar model. There will be modifications due to e.g. the spiral arms which impact local kinematics on top of the bar \citep[e.g.][]{Sellwood2019Discriminating}. The successive emergence of transient spiral arms may change the resonant orbit population, and some models consider it shaping the Hercules stream \citep{Hunt2018Transient}. This analysis is hence only a first step towards a more comprehensive model. Allowing for the additional impacts will greatly increase the number of free parameters, and estimating the bar slowing rate will necessitate making use of all available statistics, e.g. the spatial variation of the kinematic structures. Other missing factors that may affect the estimation of the bar slowing rate are: the neglected bar modes with $m > 2$, which will strengthen the minor resonances that sweeps the Solar neighbourhood before the CR; changes in bar amplitude affecting the time dependence of the resonant capturing rate; the choice of current bar pattern speed $\Omegap = 35 \kmskpc$. We note that some studies advocate a somewhat higher $\Omegap$: \cite{Sanders2019pattern} and \cite{bovy2019life} both derived $\Omegap = 41 \pm 3 \kmskpc$ by applying the continuity equation to stars in the bulge; With made-to-measure models in the bar region, \cite{Portail2017Dynamical} deduced $\Omegap = 39 \pm 3.5 \kmskpc$ and \cite{Clarke2019Milky} estimated, from fitting to proper motion data, $\Omegap = 37.5 \kmskpc$ in closer agreement with our assumed pattern speed. A higher pattern speed $\Omegap = 38 \kmskpc$ would lower our slowing rate estimate to $\eta = 0.0031 \pm 0.0008$ (using $\vphimax = 190 \kms$ to account for the shift of the CR). We note that several factors disfavour models with a higher $\Omegap$: Above $\Omegap \sim 37$, the upper separatrix of the CR would cut right through the Hercules stream. The alternative argument would be that the separatrix of the bar's CR corresponds to one of the interior structures of Hercules, e.g. separating the weak peak at $\vphi \simeq 180 \kms$ below the main clump at $\vphi \simeq 200 \kms$. However such models are inconsistent with the larger-scale structure observed, in particular with the azimuthal variation of stellar kinematics \citep[][]{Monari2019Tracing} and with the arrow-shaped structure in the $\Lz$-$\phi$ plane, which currently has no other explanation than the CR. Moreover, \cite{Binney2019Trapped} applied the Jeans’ theorem to trapped orbits in local velocity space and showed that the violation of Jean’s theorem is minimized at a bar pattern speed $\Omegap = 36 \pm 1 \Gyr^{-1} = 35.2 \pm 1.0 \kmskpc$. We note, however, that the best pattern speed that fits the Hercules stream with the bar's CR varies with the bar amplitude, the bar angle, and the underlying axisymmetric potential. Reaching consent of the best pattern speed that reproduces all observed features will require further effort.

\section{Conclusions}
\label{sec:conclusion}

While there have been extensive discussions in the literature interpreting the local velocity plane in Hipparcos and \textit{Gaia} datasets with different values of the current pattern speed $\Omegap$ of the bar, we find that the slowing rate $\dOmegap$ of the bar profoundly affects the observed substructure. Due to the highly significant and drastic impact of resonant sweeping found in this paper, we argue that any results based on a constantly rotating bar pattern speed ought to be re-examined for their robustness against this process and how their parameters have been biased by the neglect of the deceleration.

The deceleration $\dOmegap$ of a Galactic bar is a theoretical requirement resulting from the angular momentum balance of the bar: the angular momentum gain from forcing gas onto the Galactic nuclear disc is (in a standard dark matter simulation) more than offset by the dynamical friction with the dark halo (and to a minor part the surrounding disc), which implies $\dOmegap < 0$ and thus the long-term deceleration/growth of the Galactic bar. While this has been theoretically known, we are not aware of any study that would have provided a pathway to observationally estimate the long term evolution of $\Omegap$. However, by neglecting perturbations other than the bar and by investigating the effect on resonance occupation using a simple slowing bar model where the pattern speed is modeled to decline inversely proportional with time, we now provide an estimate of the current slowing rate of the bar to be $\dOmegap = (- 4.5 \pm 1.4) \kms\kpc^{-1}\Gyr^{-1}$ at current pattern speed $\Omegap = 35 \kmskpc$.

The deceleration of the bar also resolves three major issues with the appearance of the Hercules stream/corotation resonance: i) The observed Hercules stream is highly asymmetric in radial velocity $\vR$, featuring a strong outward motion. This asymmetry is underpredicted by models with a constantly rotating bar. ii) Resonant capturing by the sweeping resonance allows for larger occupation numbers than in a steadily rotating bar model, thus fitting the observed density with a reasonable bar strength. iii) The stars captured near the surface of the resonance allow for a much stronger eye-shaped (or spear-shaped for the observable Solar neighborhood) feature in the mean radial velocity $\ovR$ of the $\Lz$-$\varphi$ plane, which in the observed Solar neighborhood data explains the two strong positive $\ovR$ features near $\Lz \sim 1400$ and $\sim 1600 \kpckms$ together with their inclination against azimuth. To facilitate point (ii), we have examined how resonant capturing and retention/dragging vary with the deceleration parameter $\eta = -\dOmegap / \Omegap^2$ and the amplitude $A$ of the bar. We find that $\eta/A$ can be used as a good indicator for retention and capture and that expectations for this parameter from the observationally estimated strength $A$ and the expected slowing rate $\eta$ from N-body simulations in a typical dark matter halo place the parameter in the region, where the $\vR$ asymmetricity of the simulated local velocity plane matches that of the \textit{Gaia} data.

We stress that this work is largely of an exploratory and qualitative nature. We have not attempted to go beyond the simplest possible $m=2$ model and we have restricted ourselves to a 2D in-plane analysis. High order modes and vertical motions would bring additional resonances and more complications, which we found would have reduced the clarity of this work. We remark though that our first exploratory simulations were performed in full 3D and confirmed the same qualitative answers as presented here in 2D. Further, for the sake of simplicity, we omitted several processes that we consider to be important: spiral structure will overlay the suggested pattern, and by its transience should knock stars in and out of resonances, changing the occupation of resonant orbits. A similar role is taken by giant molecular clouds, galaxy mergers, subhalo passages, and not least, the possible jitter of the bar pattern speed itself.

We hope that this work will trigger more research into the effects of time-dependent moving resonances. A precise determination of the slowing rate of the bar from local kinematics will quantify the dynamical friction exerted on the bar and provide strong constrains on the phase-space distribution and nature of the dark matter halo.

\section*{Acknowledgements}

It is a pleasure to thank the Oxford galactic dynamics group, as well as W.Dehnen for helpful comments. RC acknowledges the support from the Takenaka Scholarship Foundation and the Royal Society grant RGF$\backslash$R1$\backslash$180095. RS is supported by a Royal Society University Research Fellowship. This work was performed using the Cambridge Service for Data Driven Discovery (CSD3), part of which is operated by the University of Cambridge Research Computing on behalf of the STFC DiRAC HPC Facility (www.dirac.ac.uk). The DiRAC component of CSD3 was funded by BEIS capital funding via STFC capital grants ST/P002307/1 and ST/R002452/1 and STFC operations grant ST/R00689X/1. DiRAC is part of the National e-Infrastructure.

\section*{Data availability}

This study used the data from \textit{Gaia} publicly available at https://gea.esac.esa.int/archive. The distances and parallax offsets for the Gaia sources are taken from \cite{Schoenrich2019distance} and are available at https://zenodo.org/record/2557803.

%%%%%%%%%%%%%%%%%%%%%%%%%%%%%%%%%%%%%%%%%%%%%%%%%%


\bibliographystyle{mnras}
\bibliography{./slowing_bar}

\begin{thebibliography}{}
\makeatletter
\relax
\def\mn@urlcharsother{\let\do\@makeother \do\$\do\&\do\#\do\^\do\_\do\%\do\~}
\def\mn@doi{\begingroup\mn@urlcharsother \@ifnextchar [ {\mn@doi@}
  {\mn@doi@[]}}
\def\mn@doi@[#1]#2{\def\@tempa{#1}\ifx\@tempa\@empty \href
  {http://dx.doi.org/#2} {doi:#2}\else \href {http://dx.doi.org/#2} {#1}\fi
  \endgroup}
\def\mn@eprint#1#2{\mn@eprint@#1:#2::\@nil}
\def\mn@eprint@arXiv#1{\href {http://arxiv.org/abs/#1} {{\tt arXiv:#1}}}
\def\mn@eprint@dblp#1{\href {http://dblp.uni-trier.de/rec/bibtex/#1.xml}
  {dblp:#1}}
\def\mn@eprint@#1:#2:#3:#4\@nil{\def\@tempa {#1}\def\@tempb {#2}\def\@tempc
  {#3}\ifx \@tempc \@empty \let \@tempc \@tempb \let \@tempb \@tempa \fi \ifx
  \@tempb \@empty \def\@tempb {arXiv}\fi \@ifundefined
  {mn@eprint@\@tempb}{\@tempb:\@tempc}{\expandafter \expandafter \csname
  mn@eprint@\@tempb\endcsname \expandafter{\@tempc}}}

\bibitem[\protect\citeauthoryear{{Alcock} et~al.,}{{Alcock}
  et~al.}{1995}]{Alcock1995Microlensing}
{Alcock} C.,  et~al., 1995, \mn@doi [\prl] {10.1103/PhysRevLett.74.2867}, \href
  {https://ui.adsabs.harvard.edu/abs/1995PhRvL..74.2867A} {74, 2867}

\bibitem[\protect\citeauthoryear{{Antoja} et~al.,}{{Antoja}
  et~al.}{2014}]{antoja2014constraints}
{Antoja} T.,  et~al., 2014, \mn@doi [\aap] {10.1051/0004-6361/201322623}, \href
  {https://ui.adsabs.harvard.edu/abs/2014A&A...563A..60A} {563, A60}

\bibitem[\protect\citeauthoryear{{Athanassoula}}{{Athanassoula}}{1992}]{athanassoula1992existence}
{Athanassoula} E.,  1992, \mn@doi [\mnras] {10.1093/mnras/259.2.345}, \href
  {https://ui.adsabs.harvard.edu/abs/1992MNRAS.259..345A} {259, 345}

\bibitem[\protect\citeauthoryear{{Athanassoula}}{{Athanassoula}}{2003}]{athanassoula2003determines}
{Athanassoula} E.,  2003, \mn@doi [\mnras] {10.1046/j.1365-8711.2003.06473.x},
  \href {https://ui.adsabs.harvard.edu/abs/2003MNRAS.341.1179A} {341, 1179}

\bibitem[\protect\citeauthoryear{{Aumer} \& {Sch{\"o}nrich}}{{Aumer} \&
  {Sch{\"o}nrich}}{2015}]{aumer2015origin}
{Aumer} M.,  {Sch{\"o}nrich} R.,  2015, \mn@doi [\mnras]
  {10.1093/mnras/stv2252}, \href
  {https://ui.adsabs.harvard.edu/abs/2015MNRAS.454.3166A} {454, 3166}

\bibitem[\protect\citeauthoryear{{Binney}}{{Binney}}{2018}]{binney2017orbital}
{Binney} J.,  2018, \mn@doi [\mnras] {10.1093/mnras/stx2835}, \href
  {https://ui.adsabs.harvard.edu/abs/2018MNRAS.474.2706B} {474, 2706}

\bibitem[\protect\citeauthoryear{{Binney}}{{Binney}}{2020}]{Binney2019Trapped}
{Binney} J.,  2020, \mn@doi [\mnras] {10.1093/mnras/staa1103}, \href
  {https://ui.adsabs.harvard.edu/abs/2020MNRAS.495..895B} {495, 895}

\bibitem[\protect\citeauthoryear{{Binney} \& {Tremaine}}{{Binney} \&
  {Tremaine}}{2008}]{binney2008galactic}
{Binney} J.,  {Tremaine} S.,  2008, {Galactic Dynamics: Second Edition}

\bibitem[\protect\citeauthoryear{{Bland-Hawthorn} \& {Cohen}}{{Bland-Hawthorn}
  \& {Cohen}}{2003}]{BlandHawthorn03}
{Bland-Hawthorn} J.,  {Cohen} M.,  2003, \mn@doi [\apj] {10.1086/344573}, \href
  {https://ui.adsabs.harvard.edu/abs/2003ApJ...582..246B} {582, 246}

\bibitem[\protect\citeauthoryear{{B{\"o}hmer} \& {Harko}}{{B{\"o}hmer} \&
  {Harko}}{2007}]{Bohmer2007BEC}
{B{\"o}hmer} C.~G.,  {Harko} T.,  2007, \mn@doi [\jcap]
  {10.1088/1475-7516/2007/06/025}, \href
  {https://ui.adsabs.harvard.edu/abs/2007JCAP...06..025B} {2007, 025}

\bibitem[\protect\citeauthoryear{{Bovy}, {Leung}, {Hunt}, {Mackereth},
  {Garc{\'\i}a-Hern{\'a}ndez}  \& {Roman-Lopes}}{{Bovy}
  et~al.}{2019}]{bovy2019life}
{Bovy} J.,  {Leung} H.~W.,  {Hunt} J. A.~S.,  {Mackereth} J.~T.,
  {Garc{\'\i}a-Hern{\'a}ndez} D.~A.,   {Roman-Lopes} A.,  2019, \mn@doi
  [\mnras] {10.1093/mnras/stz2891}, \href
  {https://ui.adsabs.harvard.edu/abs/2019MNRAS.490.4740B} {490, 4740}

\bibitem[\protect\citeauthoryear{{Buta} et~al.,}{{Buta}
  et~al.}{2015}]{Buta2015Classical}
{Buta} R.~J.,  et~al., 2015, \mn@doi [\apjs] {10.1088/0067-0049/217/2/32},
  \href {https://ui.adsabs.harvard.edu/abs/2015ApJS..217...32B} {217, 32}

\bibitem[\protect\citeauthoryear{{Chirikov}}{{Chirikov}}{1979}]{chirikov1979universal}
{Chirikov} B.~V.,  1979, \mn@doi [\physrep] {10.1016/0370-1573(79)90023-1},
  \href {https://ui.adsabs.harvard.edu/abs/1979PhR....52..263C} {52, 263}

\bibitem[\protect\citeauthoryear{{Clarke}, {Wegg}, {Gerhard}, {Smith}, {Lucas}
  \& {Wylie}}{{Clarke} et~al.}{2019}]{Clarke2019Milky}
{Clarke} J.~P.,  {Wegg} C.,  {Gerhard} O.,  {Smith} L.~C.,  {Lucas} P.~W.,
  {Wylie} S.~M.,  2019, \mn@doi [\mnras] {10.1093/mnras/stz2382}, \href
  {https://ui.adsabs.harvard.edu/abs/2019MNRAS.489.3519C} {489, 3519}

\bibitem[\protect\citeauthoryear{{Cole} \& {Binney}}{{Cole} \&
  {Binney}}{2017}]{Cole2017centrally}
{Cole} D.~R.,  {Binney} J.,  2017, \mn@doi [\mnras] {10.1093/mnras/stw2775},
  \href {https://ui.adsabs.harvard.edu/abs/2017MNRAS.465..798C} {465, 798}

\bibitem[\protect\citeauthoryear{{Contopoulos} \& {Grosbol}}{{Contopoulos} \&
  {Grosbol}}{1989}]{contopoulos1989orbits}
{Contopoulos} G.,  {Grosbol} P.,  1989, \mn@doi [\aapr] {10.1007/BF00873080},
  \href {https://ui.adsabs.harvard.edu/abs/1989A&ARv...1..261C} {1, 261}

\bibitem[\protect\citeauthoryear{{D'Onghia} \& {L. Aguerri}}{{D'Onghia} \& {L.
  Aguerri}}{2020}]{DOnghia2020Trojans}
{D'Onghia} E.,  {L. Aguerri} J.~A.,  2020, \mn@doi [\apj]
  {10.3847/1538-4357/ab6bd6}, \href
  {https://ui.adsabs.harvard.edu/abs/2020ApJ...890..117D} {890, 117}

\bibitem[\protect\citeauthoryear{{Debattista} \& {Sellwood}}{{Debattista} \&
  {Sellwood}}{2000}]{debattista2000constraints}
{Debattista} V.~P.,  {Sellwood} J.~A.,  2000, \mn@doi [\apj] {10.1086/317148},
  \href {https://ui.adsabs.harvard.edu/abs/2000ApJ...543..704D} {543, 704}

\bibitem[\protect\citeauthoryear{{Dehnen}}{{Dehnen}}{1999}]{dehnen1999simple}
{Dehnen} W.,  1999, \mn@doi [\aj] {10.1086/301010}, \href
  {https://ui.adsabs.harvard.edu/abs/1999AJ....118.1201D} {118, 1201}

\bibitem[\protect\citeauthoryear{{Dehnen}}{{Dehnen}}{2000}]{dehnen2000effect}
{Dehnen} W.,  2000, \mn@doi [\aj] {10.1086/301226}, \href
  {https://ui.adsabs.harvard.edu/abs/2000AJ....119..800D} {119, 800}

\bibitem[\protect\citeauthoryear{{Fragkoudi} et~al.,}{{Fragkoudi}
  et~al.}{2019}]{fragkoudi2019ridges}
{Fragkoudi} F.,  et~al., 2019, \mn@doi [\mnras] {10.1093/mnras/stz1875}, \href
  {https://ui.adsabs.harvard.edu/abs/2019MNRAS.488.3324F} {488, 3324}

\bibitem[\protect\citeauthoryear{{Friske} \& {Sch{\"o}nrich}}{{Friske} \&
  {Sch{\"o}nrich}}{2019}]{friske2019more}
{Friske} J. K.~S.,  {Sch{\"o}nrich} R.,  2019, \mn@doi [\mnras]
  {10.1093/mnras/stz2951}, \href
  {https://ui.adsabs.harvard.edu/abs/2019MNRAS.490.5414F} {490, 5414}

\bibitem[\protect\citeauthoryear{{Fux}}{{Fux}}{2001}]{fux2001order}
{Fux} R.,  2001, \mn@doi [\aap] {10.1051/0004-6361:20010561}, \href
  {https://ui.adsabs.harvard.edu/abs/2001A&A...373..511F} {373, 511}

\bibitem[\protect\citeauthoryear{{Gaia Collaboration} et~al.,}{{Gaia
  Collaboration} et~al.}{2018}]{GaiaDR2GaiaCollaboration}
{Gaia Collaboration} et~al., 2018, \mn@doi [\aap]
  {10.1051/0004-6361/201833051}, \href
  {http://adsabs.harvard.edu/abs/2018A%26A...616A...1G} {616, A1}

\bibitem[\protect\citeauthoryear{{Goodman}}{{Goodman}}{2000}]{Goodman2000Repulsive}
{Goodman} J.,  2000, \mn@doi [\na] {10.1016/S1384-1076(00)00015-4}, \href
  {https://ui.adsabs.harvard.edu/abs/2000NewA....5..103G} {5, 103}

\bibitem[\protect\citeauthoryear{{Gravity Collaboration} et~al.,}{{Gravity
  Collaboration} et~al.}{2019}]{Gravity2019geometric}
{Gravity Collaboration} et~al., 2019, \mn@doi [\aap]
  {10.1051/0004-6361/201935656}, \href
  {https://ui.adsabs.harvard.edu/abs/2019A&A...625L..10G} {625, L10}

\bibitem[\protect\citeauthoryear{{Grenon}}{{Grenon}}{1999}]{Grenon1999Kinematics}
{Grenon} M.,  1999, in {Spite} M.,  ed., Galaxy Evolution: Connecting the
  Distant Universe with the Local Fossil Record. p.~331

\bibitem[\protect\citeauthoryear{{Halle}, {Di Matteo}, {Haywood}  \&
  {Combes}}{{Halle} et~al.}{2018}]{Halle2018Radial}
{Halle} A.,  {Di Matteo} P.,  {Haywood} M.,   {Combes} F.,  2018, \mn@doi
  [\aap] {10.1051/0004-6361/201832603}, \href
  {https://ui.adsabs.harvard.edu/abs/2018A&A...616A..86H} {616, A86}

\bibitem[\protect\citeauthoryear{{Hernquist} \& {Weinberg}}{{Hernquist} \&
  {Weinberg}}{1992}]{hernquist1992bar}
{Hernquist} L.,  {Weinberg} M.~D.,  1992, \mn@doi [\apj] {10.1086/171975},
  \href {https://ui.adsabs.harvard.edu/abs/1992ApJ...400...80H} {400, 80}

\bibitem[\protect\citeauthoryear{{Hunt}, {Hong}, {Bovy}, {Kawata}  \&
  {Grand}}{{Hunt} et~al.}{2018}]{Hunt2018Transient}
{Hunt} J. A.~S.,  {Hong} J.,  {Bovy} J.,  {Kawata} D.,   {Grand} R. J.~J.,
  2018, \mn@doi [\mnras] {10.1093/mnras/sty2532}, \href
  {https://ui.adsabs.harvard.edu/abs/2018MNRAS.481.3794H} {481, 3794}

\bibitem[\protect\citeauthoryear{{Hunt}, {Bub}, {Bovy}, {Mackereth}, {Trick}
  \& {Kawata}}{{Hunt} et~al.}{2019}]{Hunt2019signature}
{Hunt} J. A.~S.,  {Bub} M.~W.,  {Bovy} J.,  {Mackereth} J.~T.,  {Trick} W.~H.,
   {Kawata} D.,  2019, \mn@doi [\mnras] {10.1093/mnras/stz2667}, \href
  {https://ui.adsabs.harvard.edu/abs/2019MNRAS.490.1026H} {490, 1026}

\bibitem[\protect\citeauthoryear{{Iocco}, {Pato}, {Bertone}  \&
  {Jetzer}}{{Iocco} et~al.}{2011}]{Iocco2011Dark}
{Iocco} F.,  {Pato} M.,  {Bertone} G.,   {Jetzer} P.,  2011, \mn@doi [\jcap]
  {10.1088/1475-7516/2011/11/029}, \href
  {https://ui.adsabs.harvard.edu/abs/2011JCAP...11..029I} {2011, 029}

\bibitem[\protect\citeauthoryear{{Joshi}}{{Joshi}}{2007}]{Joshi07}
{Joshi} Y.~C.,  2007, \mn@doi [\mnras] {10.1111/j.1365-2966.2007.11831.x},
  \href {https://ui.adsabs.harvard.edu/abs/2007MNRAS.378..768J} {378, 768}

\bibitem[\protect\citeauthoryear{{Kalnajs}}{{Kalnajs}}{1991}]{kalnajs1991pattern}
{Kalnajs} A.~J.,  1991, in {Sundelius} B.,  ed., Dynamics of Disc Galaxies.
  p.~323

\bibitem[\protect\citeauthoryear{{Katz} et~al.,}{{Katz}
  et~al.}{2019}]{GaiaDR2Katz2019}
{Katz} D.,  et~al., 2019, \mn@doi [\aap] {10.1051/0004-6361/201833273}, \href
  {http://adsabs.harvard.edu/abs/2019A%26A...622A.205K} {622, A205}

\bibitem[\protect\citeauthoryear{{Khoperskov}, {Di Matteo}, {Haywood},
  {G{\'o}mez}  \& {Snaith}}{{Khoperskov} et~al.}{2020}]{Khoperskov2019Escapees}
{Khoperskov} S.,  {Di Matteo} P.,  {Haywood} M.,  {G{\'o}mez} A.,   {Snaith}
  O.~N.,  2020, \mn@doi [\aap] {10.1051/0004-6361/201937188}, \href
  {https://ui.adsabs.harvard.edu/abs/2020A&A...638A.144K} {638, A144}

\bibitem[\protect\citeauthoryear{{Launhardt}, {Zylka}  \& {Mezger}}{{Launhardt}
  et~al.}{2002}]{Launhardt02}
{Launhardt} R.,  {Zylka} R.,   {Mezger} P.~G.,  2002, \mn@doi [\aap]
  {10.1051/0004-6361:20020017}, \href
  {https://ui.adsabs.harvard.edu/abs/2002A&A...384..112L} {384, 112}

\bibitem[\protect\citeauthoryear{{Lichtenberg} \& {Lieberman}}{{Lichtenberg} \&
  {Lieberman}}{1992}]{lichtenberg1992regular}
{Lichtenberg} A.,  {Lieberman} M.,  1992, {Regular and Chaotic Dynamics}

\bibitem[\protect\citeauthoryear{{Lynden-Bell} \& {Kalnajs}}{{Lynden-Bell} \&
  {Kalnajs}}{1972}]{lynden1972generating}
{Lynden-Bell} D.,  {Kalnajs} A.~J.,  1972, \mn@doi [\mnras]
  {10.1093/mnras/157.1.1}, \href
  {https://ui.adsabs.harvard.edu/abs/1972MNRAS.157....1L} {157, 1}

\bibitem[\protect\citeauthoryear{{Martinez-Valpuesta}, {Shlosman}  \&
  {Heller}}{{Martinez-Valpuesta} et~al.}{2006}]{Martinez2006Evolution}
{Martinez-Valpuesta} I.,  {Shlosman} I.,   {Heller} C.,  2006, \mn@doi [\apj]
  {10.1086/498338}, \href
  {https://ui.adsabs.harvard.edu/abs/2006ApJ...637..214M} {637, 214}

\bibitem[\protect\citeauthoryear{{McMillan}}{{McMillan}}{2017}]{McMillan17}
{McMillan} P.~J.,  2017, \mn@doi [\mnras] {10.1093/mnras/stw2759}, \href
  {https://ui.adsabs.harvard.edu/abs/2017MNRAS.465...76M} {465, 76}

\bibitem[\protect\citeauthoryear{{Minchev}, {Boily}, {Siebert}  \&
  {Bienayme}}{{Minchev} et~al.}{2010}]{minchev2010low}
{Minchev} I.,  {Boily} C.,  {Siebert} A.,   {Bienayme} O.,  2010, \mn@doi
  [\mnras] {10.1111/j.1365-2966.2010.17060.x}, \href
  {https://ui.adsabs.harvard.edu/abs/2010MNRAS.407.2122M} {407, 2122}

\bibitem[\protect\citeauthoryear{{Monari}, {Famaey}, {Siebert}, {Duchateau},
  {Lorscheider}  \& {Bienaym{\'e}}}{{Monari} et~al.}{2017a}]{monari2016staying}
{Monari} G.,  {Famaey} B.,  {Siebert} A.,  {Duchateau} A.,  {Lorscheider} T.,
  {Bienaym{\'e}} O.,  2017a, \mn@doi [\mnras] {10.1093/mnras/stw2807}, \href
  {https://ui.adsabs.harvard.edu/abs/2017MNRAS.465.1443M} {465, 1443}

\bibitem[\protect\citeauthoryear{{Monari}, {Famaey}, {Fouvry}  \&
  {Binney}}{{Monari} et~al.}{2017b}]{monari2017distribution}
{Monari} G.,  {Famaey} B.,  {Fouvry} J.-B.,   {Binney} J.,  2017b, \mn@doi
  [\mnras] {10.1093/mnras/stx1825}, \href
  {https://ui.adsabs.harvard.edu/abs/2017MNRAS.471.4314M} {471, 4314}

\bibitem[\protect\citeauthoryear{{Monari}, {Famaey}, {Siebert}, {Wegg}  \&
  {Gerhard}}{{Monari} et~al.}{2019a}]{Monari2019signatures}
{Monari} G.,  {Famaey} B.,  {Siebert} A.,  {Wegg} C.,   {Gerhard} O.,  2019a,
  \mn@doi [\aap] {10.1051/0004-6361/201834820}, \href
  {https://ui.adsabs.harvard.edu/abs/2019A&A...626A..41M} {626, A41}

\bibitem[\protect\citeauthoryear{{Monari}, {Famaey}, {Siebert}, {Bienaym{\'e}},
  {Ibata}, {Wegg}  \& {Gerhard}}{{Monari} et~al.}{2019b}]{Monari2019Tracing}
{Monari} G.,  {Famaey} B.,  {Siebert} A.,  {Bienaym{\'e}} O.,  {Ibata} R.,
  {Wegg} C.,   {Gerhard} O.,  2019b, \mn@doi [\aap]
  {10.1051/0004-6361/201936455}, \href
  {https://ui.adsabs.harvard.edu/abs/2019A&A...632A.107M} {632, A107}

\bibitem[\protect\citeauthoryear{{M{\"u}hlbauer} \& {Dehnen}}{{M{\"u}hlbauer}
  \& {Dehnen}}{2003}]{muhlbauer2003kinematic}
{M{\"u}hlbauer} G.,  {Dehnen} W.,  2003, \mn@doi [\aap]
  {10.1051/0004-6361:20030186}, \href
  {https://ui.adsabs.harvard.edu/abs/2003A&A...401..975M} {401, 975}

\bibitem[\protect\citeauthoryear{{Nakanishi} \& {Sofue}}{{Nakanishi} \&
  {Sofue}}{2006}]{Nakanishi2006ISM}
{Nakanishi} H.,  {Sofue} Y.,  2006, \mn@doi [\pasj] {10.1093/pasj/58.5.847},
  \href {https://ui.adsabs.harvard.edu/abs/2006PASJ...58..847N} {58, 847}

\bibitem[\protect\citeauthoryear{{Nidever} et~al.,}{{Nidever}
  et~al.}{2012}]{Nidever12}
{Nidever} D.~L.,  et~al., 2012, \mn@doi [\apjl] {10.1088/2041-8205/755/2/L25},
  \href {https://ui.adsabs.harvard.edu/abs/2012ApJ...755L..25N} {755, L25}

\bibitem[\protect\citeauthoryear{{P{\'e}rez-Villegas}, {Portail}, {Wegg}  \&
  {Gerhard}}{{P{\'e}rez-Villegas} et~al.}{2017}]{perez2017revisiting}
{P{\'e}rez-Villegas} A.,  {Portail} M.,  {Wegg} C.,   {Gerhard} O.,  2017,
  \mn@doi [\apjl] {10.3847/2041-8213/aa6c26}, \href
  {https://ui.adsabs.harvard.edu/abs/2017ApJ...840L...2P} {840, L2}

\bibitem[\protect\citeauthoryear{{Portail}, {Gerhard}, {Wegg}  \&
  {Ness}}{{Portail} et~al.}{2017}]{Portail2017Dynamical}
{Portail} M.,  {Gerhard} O.,  {Wegg} C.,   {Ness} M.,  2017, \mn@doi [\mnras]
  {10.1093/mnras/stw2819}, \href
  {https://ui.adsabs.harvard.edu/abs/2017MNRAS.465.1621P} {465, 1621}

\bibitem[\protect\citeauthoryear{{Regan} \& {Teuben}}{{Regan} \&
  {Teuben}}{2004}]{Regan2004BarDriven}
{Regan} M.~W.,  {Teuben} P.~J.,  2004, \mn@doi [\apj] {10.1086/380116}, \href
  {https://ui.adsabs.harvard.edu/abs/2004ApJ...600..595R} {600, 595}

\bibitem[\protect\citeauthoryear{{Reid} et~al.,}{{Reid}
  et~al.}{2019}]{Reid2019Trigonometric}
{Reid} M.~J.,  et~al., 2019, \mn@doi [\apj] {10.3847/1538-4357/ab4a11}, \href
  {https://ui.adsabs.harvard.edu/abs/2019ApJ...885..131R} {885, 131}

\bibitem[\protect\citeauthoryear{{Robin}, {Marshall}, {Schultheis}  \&
  {Reyl{\'e}}}{{Robin} et~al.}{2012}]{Robin2012Stellar}
{Robin} A.~C.,  {Marshall} D.~J.,  {Schultheis} M.,   {Reyl{\'e}} C.,  2012,
  \mn@doi [\aap] {10.1051/0004-6361/201116512}, \href
  {https://ui.adsabs.harvard.edu/abs/2012A&A...538A.106R} {538, A106}

\bibitem[\protect\citeauthoryear{{Sanders}, {Smith}  \& {Evans}}{{Sanders}
  et~al.}{2019}]{Sanders2019pattern}
{Sanders} J.~L.,  {Smith} L.,   {Evans} N.~W.,  2019, \mn@doi [\mnras]
  {10.1093/mnras/stz1827}, \href
  {https://ui.adsabs.harvard.edu/abs/2019MNRAS.488.4552S} {488, 4552}

\bibitem[\protect\citeauthoryear{{Sch{\"o}nrich}}{{Sch{\"o}nrich}}{2012}]{S12}
{Sch{\"o}nrich} R.,  2012, \mn@doi [\mnras] {10.1111/j.1365-2966.2012.21631.x},
  \href {https://ui.adsabs.harvard.edu/abs/2012MNRAS.427..274S} {427, 274}

\bibitem[\protect\citeauthoryear{{Sch{\"o}nrich}, {Binney}  \&
  {Dehnen}}{{Sch{\"o}nrich} et~al.}{2010}]{SBD}
{Sch{\"o}nrich} R.,  {Binney} J.,   {Dehnen} W.,  2010, \mn@doi [\mnras]
  {10.1111/j.1365-2966.2010.16253.x}, \href
  {https://ui.adsabs.harvard.edu/abs/2010MNRAS.403.1829S} {403, 1829}

\bibitem[\protect\citeauthoryear{{Sch{\"o}nrich}, {Aumer}  \&
  {Sale}}{{Sch{\"o}nrich} et~al.}{2015}]{Schoenrich15}
{Sch{\"o}nrich} R.,  {Aumer} M.,   {Sale} S.~E.,  2015, \mn@doi [\apjl]
  {10.1088/2041-8205/812/2/L21}, \href
  {https://ui.adsabs.harvard.edu/abs/2015ApJ...812L..21S} {812, L21}

\bibitem[\protect\citeauthoryear{{Sch{\"o}nrich}, {McMillan}  \&
  {Eyer}}{{Sch{\"o}nrich} et~al.}{2019}]{Schoenrich2019distance}
{Sch{\"o}nrich} R.,  {McMillan} P.,   {Eyer} L.,  2019, \mn@doi [\mnras]
  {10.1093/mnras/stz1451}, \href
  {https://ui.adsabs.harvard.edu/abs/2019MNRAS.tmp.1390S} {p.~1390}

\bibitem[\protect\citeauthoryear{Sellwood}{Sellwood}{2010}]{Sellwood10}
Sellwood J.~A.,  2010, \mn@doi [\mnras] {10.1111/j.1365-2966.2010.17305.x},
  409, 145

\bibitem[\protect\citeauthoryear{{Sellwood} \& {Binney}}{{Sellwood} \&
  {Binney}}{2002}]{Sellwood2002radial}
{Sellwood} J.~A.,  {Binney} J.~J.,  2002, \mn@doi [\mnras]
  {10.1046/j.1365-8711.2002.05806.x}, \href
  {https://ui.adsabs.harvard.edu/abs/2002MNRAS.336..785S} {336, 785}

\bibitem[\protect\citeauthoryear{{Sellwood} \& {Wilkinson}}{{Sellwood} \&
  {Wilkinson}}{1993}]{sellwood1993dynamics}
{Sellwood} J.~A.,  {Wilkinson} A.,  1993, \mn@doi [Reports on Progress in
  Physics] {10.1088/0034-4885/56/2/001}, \href
  {https://ui.adsabs.harvard.edu/abs/1993RPPh...56..173S} {56, 173}

\bibitem[\protect\citeauthoryear{{Sellwood}, {Trick}, {Carlberg}, {Coronado}
  \& {Rix}}{{Sellwood} et~al.}{2019}]{Sellwood2019Discriminating}
{Sellwood} J.~A.,  {Trick} W.~H.,  {Carlberg} R.~G.,  {Coronado} J.,   {Rix}
  H.-W.,  2019, \mn@doi [\mnras] {10.1093/mnras/stz140}, \href
  {https://ui.adsabs.harvard.edu/abs/2019MNRAS.484.3154S} {484, 3154}

\bibitem[\protect\citeauthoryear{{Silk} \& {Rees}}{{Silk} \&
  {Rees}}{1998}]{Silk98}
{Silk} J.,  {Rees} M.~J.,  1998, \aap, \href
  {https://ui.adsabs.harvard.edu/abs/1998A&A...331L...1S} {331, L1}

\bibitem[\protect\citeauthoryear{{Sofue}}{{Sofue}}{2013}]{Sofue2013Rotation}
{Sofue} Y.,  2013, \mn@doi [\pasj] {10.1093/pasj/65.6.118}, \href
  {https://ui.adsabs.harvard.edu/abs/2013PASJ...65..118S} {65, 118}

\bibitem[\protect\citeauthoryear{{Sormani}, {Binney}  \& {Magorrian}}{{Sormani}
  et~al.}{2015}]{sormani2015gas3}
{Sormani} M.~C.,  {Binney} J.,   {Magorrian} J.,  2015, \mn@doi [\mnras]
  {10.1093/mnras/stv2067}, \href
  {https://ui.adsabs.harvard.edu/abs/2015MNRAS.454.1818S} {454, 1818}

\bibitem[\protect\citeauthoryear{{Tremaine} \& {Weinberg}}{{Tremaine} \&
  {Weinberg}}{1984}]{Tremaine1984Dynamical}
{Tremaine} S.,  {Weinberg} M.~D.,  1984, \mn@doi [\mnras]
  {10.1093/mnras/209.4.729}, \href
  {https://ui.adsabs.harvard.edu/abs/1984MNRAS.209..729T} {209, 729}

\bibitem[\protect\citeauthoryear{{Trick}, {Fragkoudi}, {Hunt}, {Mackereth}  \&
  {White}}{{Trick} et~al.}{2019}]{Trick2019Identifying}
{Trick} W.~H.,  {Fragkoudi} F.,  {Hunt} J. A.~S.,  {Mackereth} J.~T.,   {White}
  S. D.~M.,  2019, arXiv e-prints, \href
  {https://ui.adsabs.harvard.edu/abs/2019arXiv190604786T} {p. arXiv:1906.04786}

\bibitem[\protect\citeauthoryear{{Valenzuela} \& {Klypin}}{{Valenzuela} \&
  {Klypin}}{2003}]{valenzuela2003secular}
{Valenzuela} O.,  {Klypin} A.,  2003, \mn@doi [\mnras]
  {10.1046/j.1365-8711.2003.06947.x}, \href
  {https://ui.adsabs.harvard.edu/abs/2003MNRAS.345..406V} {345, 406}

\bibitem[\protect\citeauthoryear{{Wegg}, {Gerhard}  \& {Portail}}{{Wegg}
  et~al.}{2015}]{wegg2015structure}
{Wegg} C.,  {Gerhard} O.,   {Portail} M.,  2015, \mn@doi [\mnras]
  {10.1093/mnras/stv745}, \href
  {https://ui.adsabs.harvard.edu/abs/2015MNRAS.450.4050W} {450, 4050}

\bibitem[\protect\citeauthoryear{{Weinberg}}{{Weinberg}}{1985}]{weinberg1985evolution}
{Weinberg} M.~D.,  1985, \mn@doi [\mnras] {10.1093/mnras/213.3.451}, \href
  {https://ui.adsabs.harvard.edu/abs/1985MNRAS.213..451W} {213, 451}

\bibitem[\protect\citeauthoryear{{Weinberg}}{{Weinberg}}{1994}]{weinberg1994kinematic}
{Weinberg} M.~D.,  1994, \mn@doi [\apj] {10.1086/173589}, \href
  {https://ui.adsabs.harvard.edu/abs/1994ApJ...420..597W} {420, 597}

\bibitem[\protect\citeauthoryear{{Yoshida}}{{Yoshida}}{1993}]{yoshida1993recent}
{Yoshida} H.,  1993, \mn@doi [Celestial Mechanics and Dynamical Astronomy]
  {10.1007/BF00699717}, \href
  {https://ui.adsabs.harvard.edu/abs/1993CeMDA..56...27Y} {56, 27}

\bibitem[\protect\citeauthoryear{{van Albada} \& {Sanders}}{{van Albada} \&
  {Sanders}}{1982}]{Albada82}
{van Albada} T.~S.,  {Sanders} R.~H.,  1982, \mn@doi [\mnras]
  {10.1093/mnras/201.2.303}, \href
  {https://ui.adsabs.harvard.edu/abs/1982MNRAS.201..303V} {201, 303}

\makeatother
\end{thebibliography}

%%%%%%%%%%%%%%%%%%%%%%%%%%%%%%%%%%%%%%%%%%%%%%%%%%

%%%%%%%%%%%%%%%%% APPENDICES %%%%%%%%%%%%%%%%%%%%%

\appendix

\section{Action-angles and frequencies in 2D axisymmetric potential}
\label{app:action_angle_freq_2DAxisym}

One can map from $(\vx, \vvel)$ to $(\vtheta, \vJ)$ in a 2D axisymmetric potential by numerically integrating the following equations \citep[e.g.][]{lynden1972generating}:
\begin{align}
  \JR &= \frac{1}{\pi} \int^{R_{+}}_{R_{-}} dR~p_R, ~~~ \Jphi = \pphi, ~~~ T_R = 2 \int^{R_{+}}_{R_{-}} \frac{dR}{p_R},
  \label{eq:app_actions} \\
  \Omega_R &= \frac{2 \pi}{T_R}, ~~~ \Omega_\varphi = \frac{\Delta \varphi}{T_R} = \frac{2}{T_R} \int^{R_{+}}_{R_{-}} dR ~ \frac{\Jphi}{p_R R^2},
  \label{eq:app_frequencies} \\
  \thetaR &= \OmegaR \int_C ~ \frac{dR}{p_R}, ~~~ \thetaphi = \varphi + \int_C ~ \frac{dR}{p_R} \left( \Omegaphi - \frac{\Jphi}{R^2}\right),
  \label{eq:app_angles}
\end{align}
where
\begin{equation}
  p_R(R) = \sqrt{2\left[E - \Phi_0(R)\right] - \Jphi^2 / R^2},
  \label{eq:app_pR_E}
\end{equation}
$T_R$ is the period of radial motion, and $\Delta \varphi$ is the change of azimuthal angle after one radial oscillation. The integrals in \eqref{eq:app_actions} and (\ref{eq:app_frequencies}) run from pericentre $R_{-}$ to apocentre $R_{+}$, which are the roots of
\begin{align}
  E = \Phi_0(R_{\pm}) + \frac{\Jphi^2}{2 R_{\pm}^2}.
  \label{eq:app_apsis}
\end{align}
The integration curve $C$ in \eqref{eq:app_angles} runs from the pericentre $R_{-}$ to the current radius $R$. Since the integrals in \eqref{eq:app_actions}-(\ref{eq:app_angles}) include poles at the bounds, we employ the Tanh-Sinh quadrature scheme to obtain accurate results.

The calculation of $\Psi$ and $G$ requires the inverse map from $(\vtheta, \vJ)$ to $(\vx, \vvel)$. This is not straightforward since we must find the energy given the actions. To achieve this, we precalculate the energy on a fine grid in action space $(\JR, \Jphi)$ and interpolate linearly.

\section{Fourier coefficients of the perturbing potential}
\label{app:fourier_coefficient}

The Fourier coefficients $\Psik(\vJ', t)$ in \eqref{eq:H_new} are
\begin{equation}
  \Psik(\vJ',t) = \iint^{2\pi}_0 \frac{d \vtheta'}{(2\pi)^2} \Phim(R,t) \cos\left[m \left( \varphi - \int_0^t dt'\Omegap \right) \right] \e^{- i \vk \cdot \vtheta'}. \nonumber
\end{equation}
We split $\varphi - \int_0^t dt'~\Omegap$ into $\thetaphi - \int_0^t dt'~\Omegap$ (the azimuthal angle of the guiding centre with respect to the bar) and $\varphi - \thetaphi$ (the deviation from the guiding centre which is only a function of $\thetaR$), and use \eqref{eq:slow_angle}-(\ref{eq:fast_angle}) to convert between $\vtheta$ and $\vtheta'$;
\begin{align}
  &\Psik(\vJ',t)
  = \frac{1}{2} \iint^{2\pi}_0 \frac{d \vtheta'}{(2\pi)^2} \Phim(R,t) \e^{i m \left(\thetaphi - \int_0^t dt'\Omegap + \varphi - \thetaphi \right)} \e^{- i \vk \cdot \vtheta'} \nonumber \\
  &= \frac{1}{2} \frac{1}{(2\pi)^2} \int^{2\pi}_0 d \thetas ~ \e^{i \left(\frac{m}{\Nphi} - \ks \right)\thetas} \nonumber \\
  &\hspace{2.3cm} \times \int^{2\pi}_0 d \thetaf ~ \Phim(R,t) \e^{i \left[ m \left(\varphi - \thetaphi \right) - \left(m \frac{\NR}{\Nphi} + \kf \right) \thetaf \right]} \nonumber \\
  &= \frac{1}{2}\frac{\delta_{m,\Nphi \ks}}{\pi} \int^{\pi}_0 d \thetaR \Phim(R,t) \cos \left[ m (\varphi - \thetaphi) - \left(m \frac{\NR}{\Nphi} + \kf \right) \thetaR \right], \nonumber
  \label{eq:app_fourier_coefficient}
\end{align}
where the last line is valid only for $\Nphi \le m$. $\delta$ is the Kronecker delta. $\Psi \equiv 2|\Psi_1|$ is then \citep[e.g.][]{Tremaine1984Dynamical}
\begin{equation}
  \Psi(\vJ', t) = \frac{\delta_{m\Nphi}}{\pi} \left\vert \int^{\pi}_0 d \thetaR \Phim(R,t) \cos \left[ m (\varphi - \thetaphi) - {\NR} \thetaR \right] \right\vert.
  \label{eq:app_Psi}
\end{equation}
In the limit $\JR \rightarrow 0$ at the CR $(\NR = 0)$, $\Psi = |\Phim|$.

\section{Calculation of G}
\label{app:calculate_G}

The quantity $G$ introduced in \eqref{eq:G} is
\begin{equation}
  G
  = \frac{\partial}{\partial \Js} \left(\vN \cdot \vOmega \right)
  = \vN \cdot \frac{\partial}{\partial \vJ} \left(\vN \cdot \vOmega \right) = \sum_{i,j} N_j N_i \frac{\partial \Omega_i}{\partial J_j},
  \label{eq:app_G}
\end{equation}
where the indices $i,j$ are summed over $\{R,\varphi\}$. In practice, we compute the partial derivatives of the frequencies by finite differences with $\Delta J = 1 \kpckms$. For near circular orbits ($\JR < \Delta J$), we estimate $G$ by epicycle approximation. In a logarithmic background potential, the orbital frequencies are
\begin{equation}
  \vOmega = (\OmegaR, \Omegaphi) \simeq \left(\kappa, \Omega + \frac{d \kappa}{d \Jphi}\JR \right) = \left(\sqrt{2}~, 1 - \sqrt{2} \frac{\JR}{\Jphi} \right) \Omega
  \label{eq:freq_epi_approx}
\end{equation}
and therefore $G$ is
\begin{align}
  G
  &= - \frac{\left(1 + 2 \sqrt{2} \NR / \Nphi - 2 \sqrt{2} \JR / \Jphi \right)}{\left(1 + \sqrt{2} \NR / \Nphi - \sqrt{2} \JR / \Jphi \right)^2} \frac{\Nphi^2}{\RCR^2}.
  \label{eq:G_epi}
\end{align}
At the OLR and the CR, $\JR$ is typically an order smaller than $\Jphi$ so $G$ is almost always negative. At the ILR, however, $G$ is positive.

%%%%%%%%%%%%%%%%%%%%%%%%%%%%%%%%%%%%%%%%%%%%%%%%%%

% Don't change these lines
\bsp	% typesetting comment
\label{lastpage}
\end{document}